\RequirePackage{fix-cm}
\documentclass[natbib,smallextended]{svjour3}       % onecolumn (second format)
\smartqed  % flush right qed marks, e.g. at end of proof
\usepackage{graphicx}
 \journalname{my journal}
\newcommand{\araa}{{Annual Review of Astronomy \& Astrophysics}}
\newcommand{\aap}{{Astron. Atrophies.}}
\newcommand{\apj}{{Astrophys. J.}}
\newcommand{\apjs}{{Astrophys. J. Supplement Series}}
\newcommand{\mnras}{{Monthly Notices of the Royal Astronomical Society}}
\newcommand{\nat}{{Nature}}
\newcommand{\apjl}{{Astrophys. J. Letters}}
\newcommand{\aj}{{Astronomical Journal}}

\newcommand{\apss}{{Astrophysics \& Space Science}}
\newcommand{\prd}{{Physical Review D}}
\begin{document}

\title{Gamma-ray burst progenitors %\thanks{Grants or other notes
}


\author{Andrew Levan \and Paul Crowther \and Richard de Grijs \and Norbert Langer   \and Dong Xu \and Sung-Chul Yoon   
}


\institute{Andrew Levan \at
              Department of Physics, University of Warwick, Coventry, CV4 7AL, UK \\
              \email{a.j.levan@warwick.ac.uk}           %  \\
           \and
           Paul Crowther \at
              Department of Physics and Astronomy, University of Sheffield, Sheffield S3 7RH, UK \\
              \email{Paul.crowther@sheffield.ac.uk}
              \and
              Richard de Grijs \at 
	Kavli Institute for Astronomy \& Astrophysics and Department of
Astronomy, Peking University, Yi He Yuan Lu 5, Hai Dian District,
Beijing 100871, China;
International Space Science Institute--Beijing, 1 Nanertiao,
Zhongguancun, Hai Dian District, Beijing 100190, China
\and
Norbert Langer \at 
Argelander-Institut für Astronomie der Universität Bonn, Auf dem Hügel 71, 53121, Bonn, Germany nlanger@astro.uni-bonn.de
\and
Dong Xu \at 
Dark Cosmology Centre, Niels-Bohr-Institute, University of Copenhagen, Juliane Maries Vej 30, DK-2100 København Ø, Denmark; National Astronomical Observatories, Chinese Academy of Sciences, Beijing 100012, China; Key Laboratory of Space Astronomy and Technology, National Astronomical Observatories, Chinese Academy of Sciences, 20A Datun Road, Beijing 100012, China 
\and 
Sung-Chul Yoon \at 
Department of Physics and Astronomy, Seoul National University, Gwanak-ro 1, Gwanak-gu, Seoul 151-742, Republic of Korea
}

\date{Received: date / Accepted: date}

\maketitle

\begin{abstract}
We review our current understanding of the progenitors of both long and short duration gamma-ray bursts (GRBs). 
Constraints can be derived from multiple directions, and we use three distinct strands; i) direct observations of
GRBs and their host galaxies, ii) parameters derived from modeling, both via population synthesis and
direct numerical simulation and iii) our understanding of plausible analog progenitor systems observed in the local Universe. 
From these joint constraints, we describe the likely routes that can drive massive stars to the creation of long GRBs, and our
best estimates of the scenarios that can create compact object binaries which will ultimately form short GRBs, as well 
as the associated rates of both long and short GRBs. We further discuss
how different the progenitors may be in the case of black hole engine or millisecond-magnetar models for the production of 
GRBs, and how central engines may provide a unifying theme between many classes of extremely
luminous transient, from luminous and super-luminous supernovae to long and short GRBs. 
\keywords{gamma-ray burst: general, supernovae:general}
\end{abstract}

\section{Introduction}
\label{intro}

Since their discovery in the late 1960s \citep{klebesadel73}, unveiling the origin of gamma-ray bursts (GRBs) has been a central goal of 
contemporary astrophysics. While at one point the number of proposed models was only modestly smaller than the 
number of detected GRBs \citep[e.g.][]{nemiroff}\footnote{A time of launch of the Compton Gamma-Ray observatory (GCRO) and its
BATSE instrument there were $\sim 800$ GRBs observed from a variety of missions, while
approximately 110 models had been proposed. Interestingly, despite this progress, the now favoured collapsar model 
for long GRBs was not on that list}, over the past $\sim$ 20 years we have
finally narrowed down this progenitor list. The step change in our ability to study GRBs arose from the discovery
of afterglow emission in 1997 -- these panchromatic afterglows \citep{costa,frail,vanparadijs} precisely pinpointed
GRBs on the sky, enabling their cosmological origin to be secured via observations of both afterglows and host galaxies. 
It is now clear that GRBs are exceptionally luminous cosmological explosions, with energies (if
considered isotropic) of up to $10^{54}$ erg \citep[e.g.][]{maselli14,perley14}, and redshifts ranging from
$z=0.0085$ (35 Mpc) to $z>8$ \citep{tanvir09}, possibly $z>9$ \citep{cucchiara11}. 

\begin{figure*}
  \includegraphics[width=1.0\textwidth]{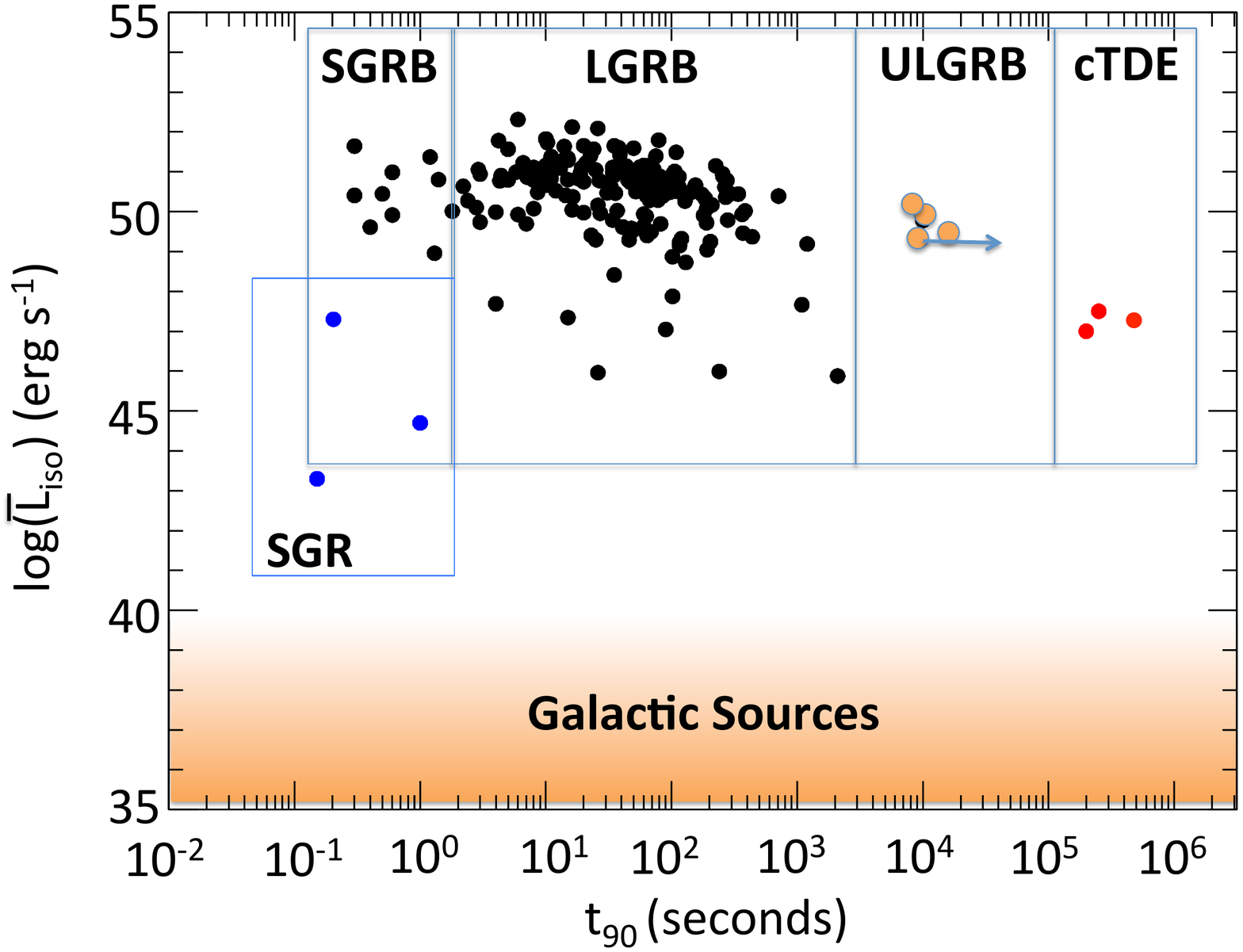}
\caption{High energy phase space for gamma-ray bursts, adapted from \cite{levan14}. The duration of the bursts is shown, compared to their
mean luminosity over that duration.  This demonstrates that
the majority of observed bursts arise from the long GRB (LGRB) or short GRB (SGRB) population, while outliers are clearly present both at extreme durations, visible
as the ultra-long GRBs (ULGRBs) and candidate tidal disruption events (TDEs) and
at low luminosities. Indeed, it is the low luminosity long GRBs that provide the best studied associated supernovae. A group of the short GRBs are also likely
to arise from giant flares from soft gamma-repeaters (SGRs) in external galaxies \citep[e.g.][]{palmer05,hurley05,tanvir05,levan08}, although these have yet to be firmly identified. }
\label{pspace}       % Give a unique label
\end{figure*}

It became apparent in the early years of GRB observation that the distribution of durations was not a smooth single population, 
but consisted of at least two peaks \citep[e.g.][]{mazets81,mazets82}. This difference, secured by observations with BATSE in the 1990s \citep{kouveliotou},
led to the identification of short  and long duration GRBs, one with a typical duration (normally defined as $t_{90}$, the time over which
90\% of the total energy release in $\gamma$-rays is recorded) of around 1\,s, and the other with a characteristic
 duration of about a minute. While further observations have identified additional possible 
sub-classes (see Figure~\ref{pspace}) at low luminosity \citep{soderberg04,soderberg06,liang}, or at intermediate \citep{mukherjee98,deugartepostigo11} 
or ultra-long duration \citep{levan11,levan14}, it is the short  and long GRBs that
make up the vast majority of the {\em observed} GRB population\footnote{The apparent distribution of GRB luminosities
is an excellent example of Malmquist bias, where the brightest events are visible over a much larger volume. Hence, while the 
observed population is dominated by high luminosity events, the volumetric rates are dominated by much lower luminosity systems
that generally escape detection}, and whose origin has been most intensely sought. 

In this review, we discuss progress towards GRB progenitors that can be made from three distinct strands. The first is
direct observations of the GRBs and the host galaxies themselves. From these observations, we can determine
the nature of any additional sources of energy in the GRB, be they supernovae (SNe) signatures in long GRBs \citep[e.g.][]{hjorth03},or radioactively
powered kilonovae (KNe) created by nucleosynthesis in the neutron-rich disc or ejecta formed 
in short GRBs \citep{barnes13,berger13,tanvir13}. These additional sources are frequently observed as late photometric bumps, interrupting the smooth
decay of the afterglow light. Their study enables information about the energy and chemical make-up of the
GRB explosion to be extracted, and has been very important in pinpointing GRBs. Studies of the afterglow light
also provide the potential to measure beaming angles in GRBs, and hence to convert the observed
rate of GRBs to the volumetric rate of GRB-like explosions. Finally, studies of the host galaxies themselves
provide information about the stellar population from which the GRB is born, and the dynamics of the progenitors. Taken
together these observations provide a significant, but still incomplete view of the stellar systems that create GRBs. 

The second strand of our consideration comes from theoretical modelling of both GRB progenitors, and the pathways that
lead to their creation. The observed GRB energetics, and the presence of photons from GRBs well above the pair production limit,
directly implies relativistic outflows \citep[e.g.][]{cavallo78}, which must somehow pierce their progenitors. This presents immediate
and significant constraints on the nature of GRB central engines that must be able to release a significant
fraction of a solar rest-mass rapidly, and into a baryon-free environment \citep[the presence of baryons would entrain
any ejecta, and make achieving relativistic velocity extremely difficult e.g.][]{lei13}. In assessing the progenitors of
GRBs from a theoretical perspective it is necessary to both model the details of the proposed progenitor (for example
its rotation, the mass of the compact remnant formed, the baryon loading in its immediate environments etc.) and
also understand the routes to obtaining these progenitors. Through this route, it is possible to determine both
the types of star that might create GRBs, and whether these systems are born at a necessary rate to match the
observed GRB population.

The third and final element of this work, and one which is often overlooked is to consider how local populations can 
inform our studies. Long GRBs are created from massive, and most likely rapidly rotating stars, and there is increasing
evidence that these are drawn from low, but not excessively low metallicity \citep[e.g.][]{fruchter06,graham13,kruhler15}. 
Indeed, most estimates have a rapid 
drop-off in the GRB rate somewhere between solar \citep{kruhler15,perley15} and 1/3 solar
\citep{graham15c} -- in other words, between the Milky Way (roughly solar) and the Small Magellanic Cloud (approximately 1/5th solar). 
In that sense, studies of the massive stellar populations within the Local Group could place strong constraints on evolutionary pathways
that are viable based on star formation in differing environments. Similarly, for short GRBs, the local population 
of compact binaries, both in the field and clusters can be used as a route to informing the rate and pathways
to their production, although again this is a challenging prospect. Such observations are likely incomplete, for example, most NS-NS binaries are found via the radio emission from a spun-up (recycled) pulsar, and there is a clear 
observational bias against binaries that merge very quickly after their initial formation (i.e. they have already
merged and cannot be observed). However, despite these
issues, local populations can provide unique diagnostics and constraints and may be able to directly
identify plausible GRB progenitors. 

By tackling the issue of GRB progenitors from these three routes it is possible to begin to constrain not 
only the basic properties of the progenitors, but also details about their likely rate, the necessary
environmental conditions for their production, the possible presence of their remnants in the local Universe, or 
the likelihood that any stars identified today may ultimately be GRB progenitors. It is also vital if we wish to
place GRBs among the full range of transients produced in the deaths of stars and to understand the links between them. 
Perhaps most import here are the links between GRBs and super-luminous supernovae \citep[SLSNe, e.g.][]{gal-yam12}. These
SNe peak a factor of 100 brighter than most SNe, but interestingly a rather similar set of models are invoked to explain their
origin \citep{gal-yam08,kasen09}, and intriguing similarities are present in their environments \citep{lunnan14a,lunnan14b,leloudas15,angus16}. 
All of this work has substantial implications
in its own right but is also a necessary first step if we  hope to use GRBs as increasingly precise cosmological tools, for
mapping the history of star formation, the build up of metals, or even as signatures of the collapse of the first stars. 

\section{What do observations tell us about long GRB progenitors?}
Progress towards understanding the nature of long-GRB progenitors has been one of the major success stories of the field and followed rapidly after the first
precise locations became available. It is now clear that at least the majority of long GRBs arise from the core collapse of massive
stars and are associated with hydrogen poor, high-velocity type Ic supernovae \citep[e.g.][]{hjorth03,cano13}. However, beyond this there
remain central questions about the progenitors that have yet to be answered; just how massive are the stars creating GRBs, and
are they classical Wolf-Rayet stars or something more exotic? 
Are binary channels important? What is the role of metallicity in creating GRB progenitors and what does this mean about
their utility as cosmic probes? Exactly what central engines are created? Are the systems creating the long GRBs similar to those
that are seen in the brightest supernovae? 

\begin{figure*}
  \includegraphics[width=1.0\textwidth]{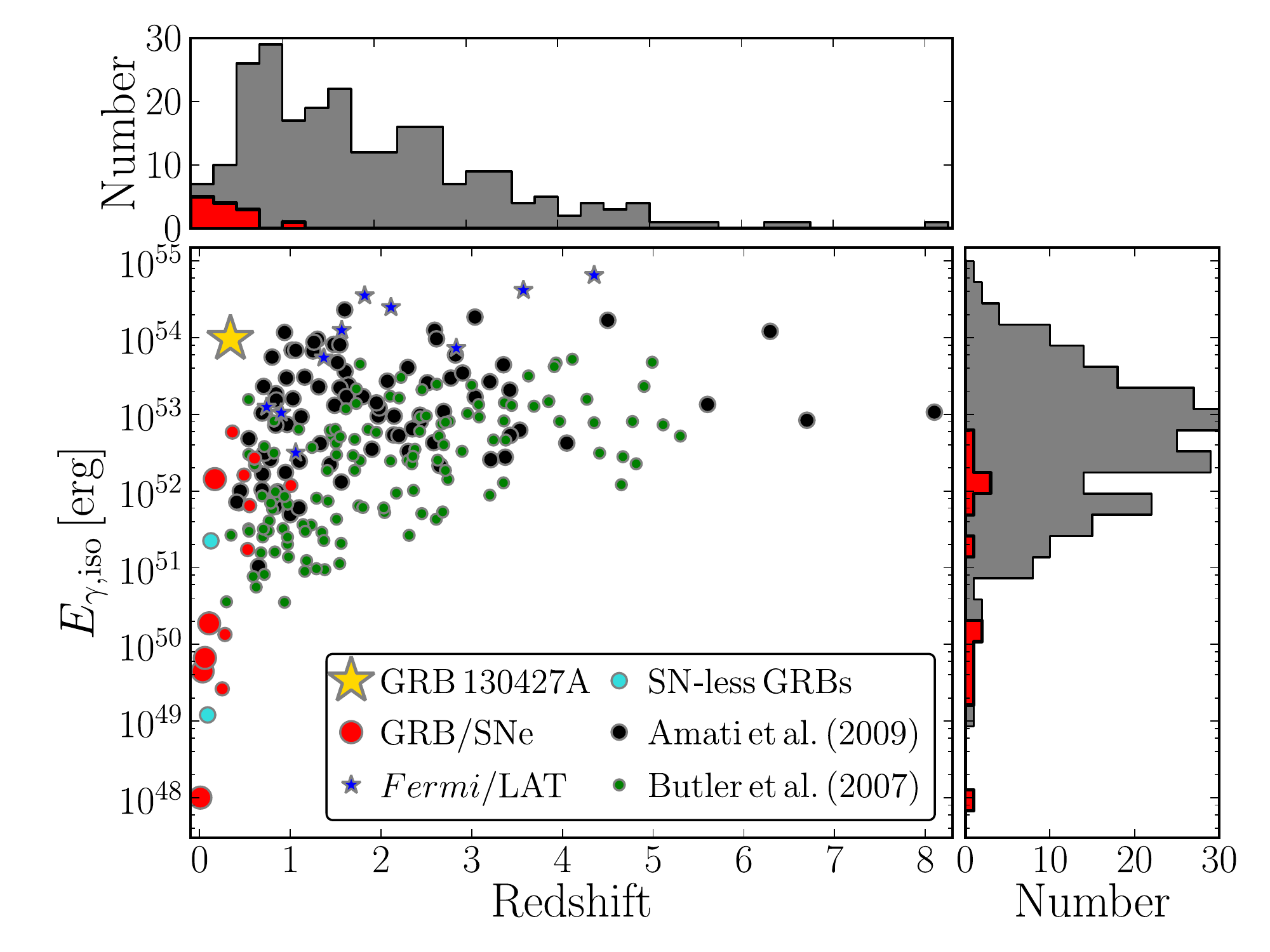}
\caption{The distribution of isotropic equivalent energy with redshift for long GRBs from \cite{xu13}, with the background
points determined by \cite{butler07} and \cite{amati09}. Those with spectroscopic detections of a supernova 
are marked in red. As can be seen these are typically low redshift events with a low luminosity (these bursts are likely intrinsically common, but
only visible at low-z). Only a handful of events are typical of the more energetic GRBs that make up the bulk of the observed GRB population, with
only GRB 130427A lying at the brightest end of the distribution.}
\label{ez}       % Give a unique label
\end{figure*}

\subsection{Building the GRB-SNe connection} 

The discovery of the extremely unusual, low luminosity GRB 980425 associated with
a broad line type Ic supernova (SN~1998bw) marked the first strong hints as to the progenitors \citep{galama98}, although the total energy
(if assumed isotropically released) of GRB~980425 was 4-6 orders of magnitude lower than seen in other GRBs, and it
had little afterglow (AG) emission \citep{kouveliotou04}. Indeed, while GRB~980425 was amongst the first handful of events
to be identified, it remains the closest known event of more than 1000 bursts with afterglows to date. 

Despite these disparate observational properties, searches for similar supernova signatures in 
more distant GRBs gradually began, and possible examples were soon found as photometric bumps, causing a re-brightening
of the optical counterpart on timescales of 20-30 days after the burst \citep[e.g.][]{bloom99}. These photometric
bumps apparently had broadly similar peak luminosities to SN~1998bw and importantly were a factor ten or more brighter
than the mean peak luminosity of core-collapse SNe. A number of examples were found in the following years, predominantly
in the GRB population observed at relatively low redshift $(z< 0.5)$ \citep[e.g.][]{price02,price03a,garnavich03,bersier06}, 
although in some cases with {\em HST} and 8-m telescopes at somewhat larger distances ($z<1$) \citep{masetti03,price03b}, and in a few
cases inferred from bumps in the absence of a redshift \citep{bloom99,gorosabel05,levan05a}. These observations gradually
built a consensus that some form of supernova was present, in at least the majority of long GRBs. However, detailed
spectroscopic study, necessary, for example, to identify if hydrogen was present, remained challenging, as the redshifts
to the majority of GRBs detected at the time lay beyond the capabilities of available spectroscopy. 

Unsurprisingly, the search was on for a low-redshift, but intrinsically energetic GRB that would enable a direct comparison
with the local, low-energy GRB~980425. Such an opportunity was presented by GRB 030329, at $z=0.17$ a redshift
at which a supernova signature could be readily visible (SN2003dh). The light curve of GRB 030329 itself was extremely complex 
\citep{lipkin04}, consisting of many re-brightenings due to ongoing energy injection \citep[e.g.][]{price03,willingale04}. However, despite
this, the optical light clearly transitioned from the featureless GRB-afterglow power-law into a spectrum with
marked similarities to that of SN~1998bw \citep{hjorth03,stanek03}, cementing the association of GRBs
with broad-lined SN Ic. 
 
\subsection{Distribution of supernova bulk properties} 
A large sample of GRB-SNe or candidate GRB-SNe has now been accrued. The majority of these arise from photometric bumps in the
late time light curve, but an increasing number also have direct spectroscopic detections of the SNe in question, although often through noisy
spectra at a single epoch\footnote{The GRB/SNe pairs with some spectroscopic evidence include GRB 980425/SN~1998bw, GRB 021211/SN~2002lt, GRB 030329/SN2003dh, GRB031203/SN2003lw, 
GRB 060218/SN2006aj, GRB100316D/SN2010bh, GRB 10129B/SN2010ma, GRB 111209A/SN~2011kl, GRB120422A/SN2012bz, GRB 130427A/SN2013cq, GRB 130215A/SN2013ez, GRB 130702A/SN~2013dx, GRB 140606B/ITPF14bfu.}. Inevitably this means that the quality of the spectra, and decisions of different authors about
the merits of including a given burst vary. However, excepting GRB 111209A/SN~2011kl \citep{levan14,greiner15} the spectra of all of these events bear marked
similarities to that of SN~1998bw, showing broad lines consistent with high-velocity expansion $>20,000$ km s$^{-1}$, and no sign of hydrogen or helium emission
features. All of these SNe are therefore spectroscopically classified as broad-lined type Ic supernovae (SN Ic-BL). 

The task of obtaining broad-band 
spectral shapes and peak luminosities from light curves of these SNe is made challenging by their large luminosity distances, 
by the contribution of the host galaxy, and by the afterglow contribution. Indeed, at any given time the observed flux depends on the 
contribution of all three, e.g. 

\begin{equation}
F_{\mathrm{obs}} (\nu,t) = F_{\mathrm{AG}} (\nu, t) + F_{\mathrm{SN}} (\nu, t) + F_{\mathrm{host}} (\nu).
\end{equation}

$F_{\mathrm{host}}$ can in principle be obtained from late-time observations and subtraction from early data, although in practice this is often complicated
by slightly different filters or instrument combinations, as well as the difficulties introduced by matching ground-based seeing. A typical GRB 
host is of comparable (or perhaps slightly lower) luminosity than its SNe \citep{savaglio09,svensson10,perley15} and so can create a significant
uncertainty, especially for higher redshift bursts, and even more so when spectroscopy is used (since the spectroscopy is inevitably noisier
than imaging). The afterglow can normally be treated as a power-law in both frequency (or wavelength) and time
and this approach is normally used e.g.,

\begin{equation}
F_{\mathrm{AG}} (\nu, t) = t^{-\alpha} \nu^{-\beta}.
\end{equation} 

Measurements of the afterglow can be complicated by unseen temporal breaks (e.g. the jet-break due to lateral spreading
of the GRB-jet, which typically steepens $\alpha$ to $\sim 2$, \citep[e.g.][]{rhoads99}) or spectral breaks (e.g. the cooling break,
which imparts $\Delta \beta = 0.5$ \citep[e.g.][]{sari98}). 

Given the spectral similarities to SN~1998bw, a common approach is to assume that the light curve of the SN evolves in 
time, $t$, like that
of SN~1998bw but modified by a stretch parameter $s$, and luminosity scaling\footnote{Here we use $h$ to determine the luminosity scaling 
to avoid confusion with the k-correction, however in other works it is common to see the luminosity scaling expressed as $k$}, $h$. 

\begin{equation}
F_{\mathrm{SN}} = h F_{\mathrm{98bw}} (t/s),
\end{equation}

where $F_{\mathrm{98bw}}$ is the specific flux of SN~1998bw as observed at the same rest-frame wavelength and time as that of the burst under consideration (i.e.
taking into account a k-correction, and cosmological time dilation). 

One can fit for these things simultaneously on well-sampled data \citep[e.g.][]{zeh04,kann,cano13,cano14}, although degeneracies can exist (e.g. between $h$ and $\alpha$), and it
is also common to attempt to extrapolate the afterglow contribution, either from earlier observations, or from data in the UV or IR where
the SNe contribution is small \citep[e.g.][]{levan14b}. 

In practice, this is a relatively crude approach, since the SNe light curve may not be a good approximation to a stretched and scaled SN~1998bw. However, in
many cases it appears to work well, at least given the data available. More recent attempts have moved beyond simple scaling factors and have attempted
to infer bolometric properties, either from the scaled SN~1998bw light curves \citep[e.g.][]{cano13} or using other SNe as templates, enabling single colours
to be used for bolometric corrections \citep[e.g.][]{lyman14b,lyman16a}.

One thing which is clearly apparent in these observations is that the classical long duration, highly energetic GRBs that comprise the bulk of the {\em observed}
GRB population are highly under-represented in the GRB-SNe sample (Figure~\ref{ez}). In itself, this is not surprising, since SNe are most readily seen in low
redshift examples, and when not outshone by the afterglow (whose luminosity broadly scales with that of the $\gamma$-rays, \citep[e.g.][]{nysewander,gehrels08}). Indeed,
many of the GRBs arise from a potentially distinct population of low-luminosity GRB (LLGRB), while even the Rosetta Stone of GRB 030329/SN~2003dh is
of rather intermediate luminosity. Only GRB~130427A/SN~2013cq appears to arise from a highly luminous GRB \citep[][]{xu13,levan14b,melandri14}. 

While this may not be problematic, it is notable that in some cases these local, low luminosity bursts appear very different from their higher energy
cousins. In particular, some are extremely long and dominated by thermal emission \citep{campana06,starling11}, features often not seen in 
other GRBs. It this sense it may be that they are not indicative of GRB-SNe in general, although it may also be that similar components in
more distant, luminous GRBs escape detection, and some thermal components have been found in careful searches \citep{starling12,sparre12}, perhaps suggesting
similarities. Indeed, the strongest link of the similarities between these very different energies of high energy transient is actually the properties
of their supernovae, which are sufficiently similar that it is likely they all arise from the same physical mechanism. 

\subsection{Test-bed examples}
An alternative to the large scale samples accrued with data of variable quality is to attempt detailed modelling of well-studied examples, where
time series spectra and excellent photometric coverage is available. In these cases it is possible to go beyond simple analytical fits to light curves, 
or snapshot velocities from spectroscopy, and build detailed models of the explosions of a range of stars that map the nickel releases, velocities, and kinetic energies, providing
high-quality spectrophotometric predictions that can be compared with observations. The principle here
is to conduct detailed spectral synthesis, in which energy injection from the core is coupled with radiation transport to predict the observed spectrum at
given epochs as a function of the various input parameters. In principle, these models can also include detailed geometry, for example, ejecta which
includes anisotropies in order to match both the rise/decay times of the SNe. This avoids the need to use scaling relations to obtain SNe properties and
instead obtains them from ab initio approaches. It has been successful in several GRBs, including GRB 980425/SN~1998bw \citep[e.g.][]{hoflich99,woosley99,maeda02}, GRB 030329/SN~2003dh \citep[e.g.][]{deng05} and GRB 060218/SN~2006aj \citep{pian06,mazzali06}. Results from such approaches are reassuringly often
similar to those from light curves alone in terms of bulk properties \citep[e.g. nickel mass,][]{cano13}, although also provide evidence at times
for anisotropies in the explosion, and stronger constraints on pre-explosion core mass and ejecta mass. However, the observational requirements for such work are frequently extreme, and so for the majority of GRBs such detailed information is
either too expensive or impossible to obtain via current instrumentation. 

\subsection{Pushing the boundaries: from no supernovae to the most luminous supernovae}
While broad-lined type Ic supernovae have been discovered in the vast majority of cases where such a search was plausible there are a handful of cases where
such searches have been unsuccessful. In particular, two local long GRBs, 060505 and 060614 (at $z=0.09$ and $z=0.125$) lie at redshifts where
 SNe similar to SN~1998bw should be readily identified, with peak apparent magnitudes in the range $18.5 < m_V < 19.5$. However, deep searches
 failed to identify any such signatures to limits not only much fainter than GRB-SNe but also to other core-collapse events. Indeed, any SNe in these cases
 must have been a factor $>100$ fainter than SN~1998bw \citep{dellavalle06,fynbo06,gal-yam06,gehrels06}. It has been noted that GRB 060505 was 
 of relatively short duration $\sim 4s$, while GRB 060614 has subsequently been suggested to be an example of a short burst with extended, softer emission 
 \citep{gehrels06}. This scenario seems likely, since further examples of SN-less, but unambiguously long GRBs have not been uncovered.  However, other suggestions, for
 example, that these GRBs may arise in cases where no outward supernova shock is launched and the star collapses directly to a black hole \citep{fynbo06,fryer07} have
 also been made. This may be particularly relevant given the apparent absence of very massive progenitors to local SN II-P \citep{smartt09}, which 
 has led to renewed interest in the prospect of disappearing stars \citep{kochanek,kochanek2}, with recent surveys beginning to find candidate examples \citep{reynolds}. 
 Surprisingly, despite the decade since these discoveries, further examples have not been found, and so their nature remains mysterious. 
 
 At the other end of the scale, the recent discovery of an SN-GRB (SN~2011kl/GRB 111209A) which was a magnitude more luminous than SN~1998bw \citep{greiner15,kann16}, challenges the picture of 
 SN homogeneity that has been emerging from previous observations. This SN was found in an ultralong GRB with a duration of $>10,000$s \citep{gendre13,levan14}, and
 in addition to being more luminous than most GRB-SNe was also spectrally very different. In particular, while it also appears to belong to the SN Ic population, it
 was far bluer and more UV luminous than other GRB-SNe, and its spectrum bore a strong resemblance to the spectra of type I SLSNe, which also show weak
 absorption lines on an extremely blue continuum \citep[e.g.][]{mazzali16}. Other ULGRBs do not show such strong apparent SNe, and so it remains unclear if
  this burst is a unique and unusual object, or if it in fact represents a broader range of SNe properties that should be considered, perhaps from cases
where the central engine of the GRB begins to impact the supernova itself, as is likely the case for SLSNe \citep[e.g.][]{cano16}. 

\subsection{Large scale environments}
Additional constraints on the progenitors of long GRBs can be obtained from both their large scale and small scale environments. The metallicities of their
host galaxies are of particular interest (in practice, the metallicities of the small scale environments are also of interest, but are not readily accessible with 
current technology, since one arc second seeing typically corresponds to several kiloparsecs at the redshift of a typical long GRB). 
Such constraints can either come from direct spectroscopic observations, or via inferences using the well-known correlation between
mass (or luminosity) and metallicity. GRBs can provide exceptional diagnostics of metallicity at high redshift via studies of UV absorption lines in concert with
the direct detection of Ly$\alpha$ \citep[e.g.][]{fynbo09}. However, this is predominantly for higher redshift bursts, since Ly$\alpha$ does not enter the
optical window until $z>2$, and a measurement of the hydrogen column density $N_{\mathrm{H}}$ along the line of sight is necessary to obtain the
ratio of a given element to hydrogen (e.g. O/H, Fe/H). For this reason the metallicities of the host galaxies where the GRB-SNe connection has been determined are
predominantly obtained either from emission line diagnostics such as, $R_{23}$  = ([O{\sc ii}] (3727\AA) + [O{\sc iii}] (4959+5007\AA)) / H$\beta$ or N2  = $\log$ [N{\sc ii}] / H$\alpha$, or indeed more complex approaches to line ratios that attempt to remove degeneracies that exist is the more simple examples \citep[e.g.][]{dopita16}. 
This provides some insight into the chemical state of the gas-phase in these galaxies, although several important caveats should be noted in the
use of these metallicities as a direct insight into GRB progenitors. Firstly, they are not a measurement of the metallicity of the progenitor star, and indeed, 
even if they are an accurate measurement of the oxygen metallicity of the star, this is still some way from the [Fe/H] ratios that are more commonly used
in distinguishing different stellar evolution pathways. Secondly, significant metallicity gradients can exist within galaxies, and so the use of global proxies 
essentially derives central metallicities (i.e. those of the brightest regions), which may, or may not be indicative of the regions hosting a given transient event. Thirdly, in some cases, the GRB may be associated with a satellite galaxy of the presumed host. For example, \cite{kelly13} show that GRB 130702A (at $z$=0.145), lies in a satellite galaxy at a significant
offset (18 kpc). However, in other cases it is likely that more proximate galaxy/satellite pairs are not adequately resolved, at least by ground-based imaging, leading
to the misidentification of the host. In extremum, there can even be chance alignment in which a background galaxy aligns with a foreground system leading
not only to the misidentification of the precise host galaxy but the incorrect determination of the redshift. This has recently been demonstrated for
an extremely well studied optically dark (i.e. no optical afterglow) GRB 080219B, which, rather than lying at $z = 0.41$ is in fact in a background galaxy at $z=1.96$ \citep{perley16x}.
In these cases, the use of metallicity measurements inferred from the ``apparent" host is clearly incorrect, and potentially problematic. For example, the
case of GRB 080219B has been used to argue for the presence of GRBs in high metallicity environments. but in practice provides no such evidence since this
metallicity is simply of the foreground system. 

\begin{figure*}
  \includegraphics[width=0.7\textwidth,angle=270]{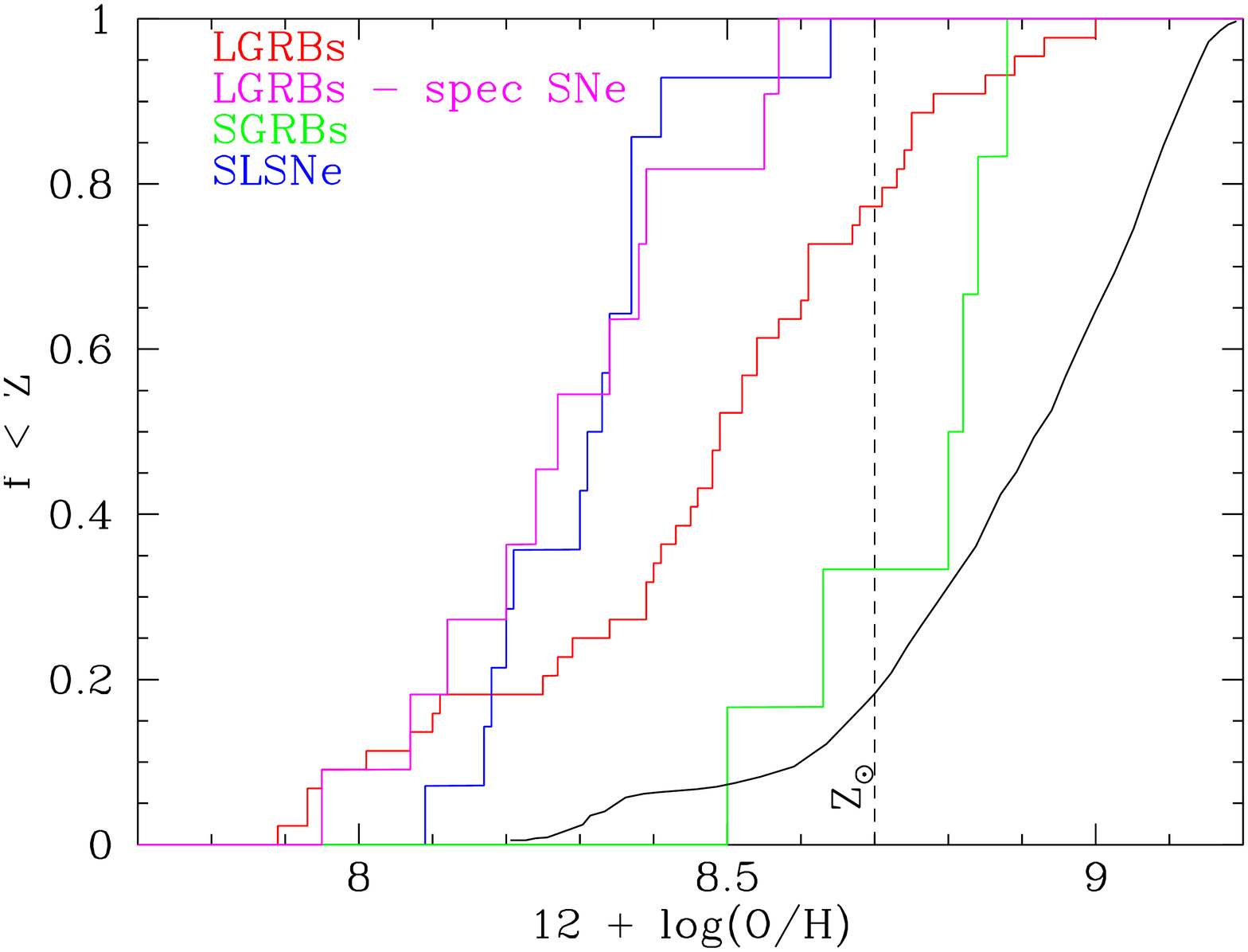}
\caption{Host galaxy metallicities for long GRBs from \cite{kruhler15}, short GRBs \cite[from][]{berger09}) and SLSNe \cite[from][]{leloudas15}. It is striking
that the metallicities of LGRBs appear significantly lower than for the SGRBs, or indeed the general galaxy population \citep{graham13}. However, those
of the GRBs with apparent SNe appear to be somewhat lower still, and are very similar to those observed in SLSNe host galaxies. }
\label{met}       % Give a unique label
\end{figure*}

The range of metallicities seen in long GRB host galaxies (as well as short GRBs and luminous SNe) is shown in Figure~\ref{met}. It is clear that GRBs favour
a lower metallicity than is typical in the local Universe, although given the redshift range considered this is not surprising. Interestingly, the metallicities
of long GRB hosts at large appear systematically higher than those of the examples in which a spectroscopic signature of the SNe has been seen, with the
median metallicity of the GRB-SNe sample being 0.2-0.3 dex lower. This may well be due to a bias against dusty systems, since obscured star formation is
typically more metal rich, and it is clear that dusty GRBs arise in more luminous, and likely metal-rich host galaxies. In any case, it also interesting to note, 
that when considering only the GRBs in which spectroscopic SNe signatures are seen there is an apparent similarity between their metallicity distribution and those of SLSNe. The degree to which these differences arise due to observational selection effects, or alternatively due to genuine astrophysical 
differences is of significant interest for further study. It may reflect genuine difference between populations, or perhaps that even in the relatively low
redshift regime there may be significant selection effects \citep[e.g.][]{vergani16}

\subsection{Small scale environments} 
\label{lgrb_loc}
The large scale environments can provide information about the chemical state of GRB hosts that is not currently available at smaller scales, due to both large
luminosity distance, and the inability of large aperture telescopes to probe with sufficiently high resolution. However, star formation takes place on small
spatial scales, from less than a parsec to tens of parsecs depending on the mode and intensity of star formation \citep[e.g.][]{zwart10}. It is these small scale environments that in practice carry most information about the GRB progenitor itself, free from dilution from the light of the
remaining galaxy. At a typical GRB redshift, the {\em HST} resolution (0.1 arcsec) still corresponds to physical scales of several hundred parsecs, and so 
while we are able to track the stellar populations in the regions of the GRBs, isolating the stellar population responsible for the GRB remains challenging. 
Two approaches can be considered to studying GRB locations, the first is simply to identify the GRB location relative to its host galaxy by measuring its offset, either
relative to the galaxy host \citep[e.g.][]{bloom02}, or perhaps to local regions of intense star formation, if such resolution is available \citep[e.g.][]{hammer06}. The alternative
is to try to study the population under the burst, despite the poor spatial resolution. This may be highly diagnostic for long GRBs, since the young massive
stars are by far the most luminous in the host galaxy, especially at UV wavelengths \citep{fruchter06}. Indeed, at the typical redshifts of many GRBs 
rest frame UV light is redshifted into the optical window, making sensitive high-resolution imagery with {\em HST} an ideal route to characterising the
immediate massive star populations surrounding GRBs. In addition, these hot massive stars  
also excite the gas phase of the interstellar medium, making
narrow-band observations a sensitive probe of the young massive star population \citep{james06}. 

Such an approach has been taken by various groups studying both supernovae and GRBs. \cite{fruchter06} consider the total fraction of the galaxy light
in pixels of lower surface brightness than the pixel containing the GRB or SNe, the so-called $F_{\mathrm{light}}$ parameter. They show that GRBs are highly 
concentrated on the light of their host galaxy, while core collapse SNe broadly trace the distribution of rest-frame UV-light. In other words, the
probability of a SN occurring in a given pixel is approximately proportional to the brightness of that pixel, in contrast, the probability of a GRB
occurring in a given pixel is proportional to the brightness of that pixel squared\footnote{The squared exponent here is approximate, and does not carry any
specific meaning}. Interestingly, further studies of local SNe show that a similar effect can be seen in the distribution of hydrogen-rich SNe II and hydrogen/helium poor SNe Ic
\citep{james06,kelly08,anderson12}. This is likely understood because
of the strong correlation between stellar mass and luminosity. In the low mass range (10-20 M$_{\odot}$) $L \sim M^3$, while for higher mass stars (50-100 M$_{\odot}$) $L \sim M^2$ 
\citep{yusof13} such that the most massive stars dominate the UV budget. In this sense, if
core collapse SNe arise from essentially every star with an initial mass $>8$M$_{\odot}$, then GRBs must arise from stars significantly more massive
than this. Studies based on the expected distribution of stars in galaxies \citep{raskin08}, and on young clusters in nearby galaxies \citep{larsson07} provide
a consistent picture in which GRB progenitors have initial masses of $>40$ M$_{\odot}$, representing one of the few ways in which the masses
of GRB progenitors can be ascertained.

\subsection{Summary: LGRB progenitors} 

The emerging picture for long GRB progenitors appears clear. A long GRB progenitor is a massive star, typically born
at sub-solar metallicity, in which the hydrogen envelope is lost/burned prior to its explosion as a broad-lined
type Ic supernova. There is some observational evidence that suggests these stars are very massive, for example, the large Nickel yields required to achieve absolute magnitudes a factor 10 brighter than SNe Ic \citep[e.g.][]{mazzali07b}, or the location of the
bursts on the brightest regions of their hosts \citep{fruchter06,svensson10} could both be interpreted as signatures of very massive stars, perhaps with with an initial mass at the zero-age main sequence (ZAMS) of $M_{\mathrm{ZAMS}} > 40$M$_{\odot}$. However, the locations of long GRBs are similar to those of the bulk SNe Ic population \citep{kelly08}, most of which can
be explained by the explosion of initially far less massive stars \citep[e.g.][]{mazzali07,mazzali07c}, and more complex models, involving the impact of chemically homogeneous evolution \citep[e.g.][]{szecsi15}, or
binaries \citep[e.g.][]{stanway16} on the observed environments may be necessary to provide tighter constraints on the progenitor mass, as well as on the progenitor rotation. These scenarios are not
currently well constrained observationally, and theoretical progress is discussed in section~\ref{theory}. None-the-less, the progress which
has been made towards the progenitors of long GRBs has been remarkable, and while there remain many open questions about the details of long GRB
progenitors, it is fair to say that barring a few special examples, we now know what forms the vast majority of the long GRB population.
\section{Observational constraints on short GRB progenitors}

\subsection{Defining short GRBs}

Short GRBs are traditionally defined as those with durations of $t_{90} <2 s$. Observations with {\em Konus} and {\em BATSE} suggested they form a separate population from 
the long GRBs, and so may well also have different progenitors \citep{mazets81,mazets82,kouveliotou}. Following the identification of supernovae signatures in 
the afterglows of long GRBs, the favoured model for short bursts rapidly became the merger of two compact objects
(neutron star -- neutron star or neutron star -- black hole), a model which had previously been popular for the long GRB population \citep[e.g.][]{narayan92}. 
The reason for this was two-fold; both that the rapid time scale for the merger, combined with the clean environment (compared to
e.g a collapsing massive star) meant that short durations were more naturally expected in merger scenarios. The second
was the rather more speculative assertion that if mergers did not produce the long bursts, they must either produce the short bursts, or
no high energy transient at all. Given this historical approach, it is unsurprising that the observational history of short GRBs 
has largely been benchmarked against this expectation, and hence the discussion of short GRBs has been largely framed as though their
progenitors are known, even prior to the accrual of the significant observational data that now supports a merger origin.

\begin{figure*}
  \includegraphics[width=0.7\textwidth,angle=270]{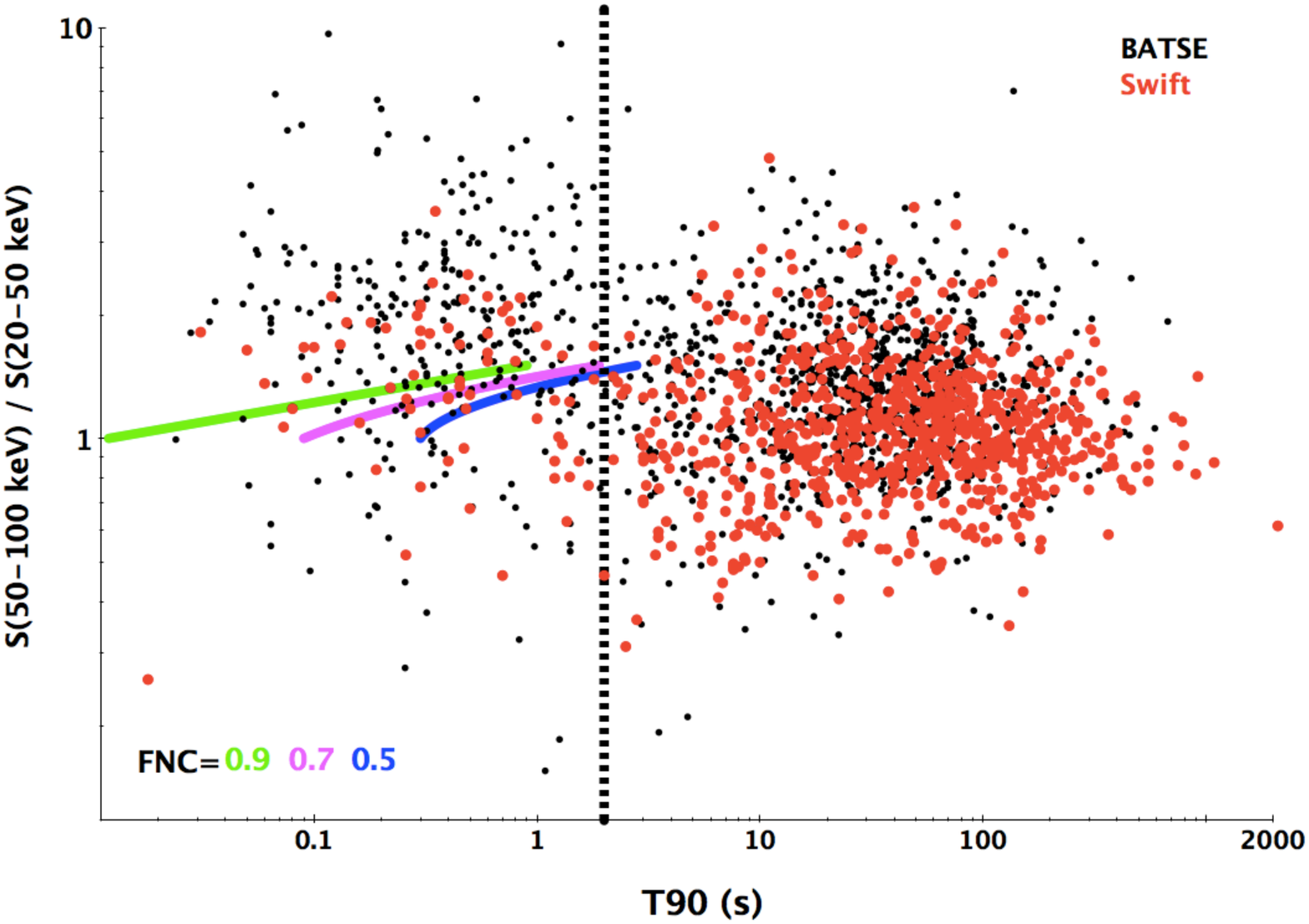}
\caption{The hardness-duration distribution of GRBs detected by {\em Swift} (red) and BATSE (black). The x-axis shows the duration over which 90\%
of the total fluence is recorded ($T_{90}$, while the y-axis shows the $\gamma-$ray hardness, defined as the ratio of flux between two adjacent channels
(the channels chosen are those used by BATSE). The {\em Swift} data have been placed on the same scale by extrapolating the prompt spectral fits (either
as a power-law or cut-off power-law) to the appropriate bands. The dotted vertical line shows the notional distinction between short and long GRBs at two seconds, but
there is clearly significant overlap. It is apparent from the BATSE distribution that short GRBs are {\em on average} harder than long GRBs. However, for {\em Swift} the
two distributions look very similar. Given this the contribution of the long GRB population is larger for {\em Swift} bursts at durations of $<$2s than for BATSE. The 
lines suggested by \cite{bromberg} as distinguishing long and short GRBs at 50,70 and 90\% confidence are also shown (where FNC is the fraction
of non-collapsars), bursts below these lines are more likely to be collapsars from the short duration tail of the long GRB population. For {\em Swift}, only bursts above the 90\% green line
are highly likely to be genuine members of the short burst population. }
\label{grb_hd}       % Give a unique label
\end{figure*}

Short bursts now represent approximately 10\% of the bursts observed by {\em Swift}, down somewhat from the 25\% 
observed in the BATSE observations. It seems likely that a significant element of this is due to the differing 
spectral responses of the different instruments, with BAT operating in a rather softer 15--150 keV band compared to BATSE (50--$>$300 keV). 
This has led to some discussion of whether the duration split between the two populations is correctly set at 2 seconds. Indeed, the
duration distribution of {\em Swift} GRBs is far less clearly bimodal than that for BATSE, and so one
cannot directly distinguish a ``cut-off" based on these observations alone. Several alternative attempts in the early years of 
{\em Swift} addressed the potential problem by suggesting the use of additional diagnostics to distinguish between 
different progenitor types \citep[e.g.][]{levan07,zhang09}. Indeed, \cite{zhang09} argued for a distinction between 
Type I (compact object merger) and Type II (collapsar) events based on a decision tree of multiple observation properties. These additional constraints often used 
rather more complex properties (for example the nature of the host galaxy, or the presence/absence of a supernova component) and so also
risked significant confirmation bias, in finding bursts which met the expectations based on only a handful of systems and omitting
other, potentially valuable systems.  Using theoretical considerations, largely based on the
break-out time for a collapsar, combined with the different spectral responses of different observatories \cite{bromberg13} have
suggested that very different values for $T_{90}$ would be needed to identify the split between long  and short GRBs for different missions. 
In particular, 2s for BATSE and 0.7 s for {\em Swift}\footnote{At 50\% probability. Since the standard assumption is to think of the duration distribution of
long  and short GRBs as Gaussians, there is inevitably overlap in the two populations}. The essence of the argument is that
hardness is a more important distinction than duration at durations where there is likely to be significant overlap between
the two populations. This is shown graphically in Figure~\ref{grb_hd}, where the {\em Swift} short GRBs can be seen to be typically
much softer than those detected by BATSE. Applying a cut of this nature would remove some (although by no means all) of the best studied 
``short GRBs" from consideration, since they would then be more likely to be collapsars.

\subsection{Precise locations from afterglows}

While progress towards the origin of long GRBs proceeded at pace from the discovery of the first afterglows to the identification of 
broad-lined SNe-Ic, pinpointing the progenitors of the short GRBs remained much more challenging. Firstly, 
short GRBs are typically significant fainter, or more precisely exhibit markedly lower fluence than long GRBs despite similar 
peak fluxes (a parameter which more accurately describes their detectability to many $\gamma$-ray observatories). Since
broad correlations exist between the fluence of the prompt emission and the brightness of either the X-ray or optical afterglow
\citep{gehrels08,nysewander09} it is perhaps unsurprising that the afterglows of short GRBs escaped detection while the afterglow
revolution was transforming our knowledge of the long GRB phenomena. Indeed, while a handful of short GRBs were 
detected and reasonably localised during the long GRB afterglow revolution, no successful afterglow campaigns were 
made prior to the launch of {\em Swift}. Its ability to re-point its X-ray and UV-optical telescopes rapidly provided
the first X-ray afterglow detection for GRB 050509B \citep{gehrels05,hjorth05a,bloom06}. Interestingly, the first
optical afterglow detection actually came from a HETE-2 burst, GRB 050709 \citep{hjorth05b,fox05}, which with the
detection of another HETE-2 burst in early 2006 \citep{deugartepostigo06,levan06}, 
perhaps suggested that previous searches had been rather unlucky, although doubtless the ability of {\em Swift} to
rapidly repoint and detect their X-ray afterglows was also an important component in the discovery of their optical
counterparts.

As with long GRBs, the discovery of afterglows to short GRBs has revolutionised their study. 
Perhaps the most important diagnostic enabled by an afterglow is a precise location on the sky, and hence
the ability to study the galaxy population hosting short GRBs. 
The first short burst with an afterglow was GRB 050509B, and its location was striking, being
offset $\sim 30$ kpc from an extremely massive galaxy in a merging
cluster system at $z=0.225$ \citep{gehrels05,bloom06}. Deep searches for star formation in this 
host galaxy revealed no sign, and immediately suggest that some short GRBs arise from ancient populations. 
However, this burst was localised only by its X-ray afterglow, and the large resulting error box contained
additional background galaxies, many of which had blue colours consistent with star formation. 
Hence, while the probability of chance alignment was low, the association did not clinch the origin 
of at least some short GRBs in ancient populations. The discovery of the optical afterglow of GRB 050724 in a clearly
elliptical host galaxy further strengthened this argument \citep{berger05}, and while bursts in low star formation rate
host galaxies are clearly in a minority, perhaps 10-20\% of short GRBs overall do arise from elliptical hosts \citep{fong13}. 
Equally, while the remaining 80-90\% of the host galaxies do show signs of significant star formation, they remain
distinct from the hosts of the long bursts in terms of their stellar masses and star formation rates. While long GRBs
show a strong preference for star-forming dwarfs, with low metallicities and high specific star formation rates
\citep[e.g.][]{fruchter06,savaglio09,svensson10,graham13}, the short bursts are in rather more typical galaxies, with
a range of metallicities and higher stellar masses, therefore they appear to sample the entirety of the galaxy
population \citep{leibler10,fong13}. 

Beyond the nature of the galaxies themselves, there is significant information contained within the
distribution of the short GRBs on their hosts. Again, the differences between long  and short GRBs become 
apparent rapidly in this distribution. While long GRBs are highly concentrated on their host light \citep[][see section~\ref{lgrb_loc}]{bloom02,fruchter06,svensson10}
the short GRBs are frequently scattered on their host light, and at large projected radii, in some cases there
is apparently no underlying host galaxy, despite relatively bright galaxies nearby in the field, suggesting that
these progenitors have been kicked from their birth sites \citep{berger10,tunnicliffe14}, although it remains
plausible that some lie at much higher redshifts in galaxies too faint for current observations \citep{levan06,berger07}. 
Beyond this much broader distribution, those short GRBs that do lie on galaxies tend to appear on much fainter pixels
than those in the long GRBs \citep{fong10,fong13b}, and in particular show essentially no association with the blue
light of their host galaxies. Taken together these properties demonstrate that short GRB progenitors are frequently old, and
more importantly are also kicked from their birthplaces. These are all consistent with the expectations 
of binary mergers, in which the neutron stars receive substantial space velocities (kicks) from
 a combination of natal kicks and binary mass loss \citep[e.g.][]{arzoumanian02}.

\subsection{Kilonovae and macronovae}
The large scale observations are clearly broadly in favour of a model in which short GRBs arise from
compact object mergers. However, they are in themselves not a smoking gun of the merger itself. This
situation is reminiscent of early studies of long GRBs in which the star-forming host galaxies offered evidence
of a link to massive stars and supernovae, but did not conclusively prove it. Since afterglows are generated at large radii and are
essentially a product of the interaction of an outgoing relativistic shock with an external medium, they themselves
are not uniquely diagnostic of the progenitor. Hence, some additional signature that could in principle only
arise in compact object mergers was necessary. The most likely scenario is to identify a faint radioactively
powered transient, created by nucleosynthesis in the neutron-rich material available in a compact object
merger containing a neutron star. This material may reside either in the accretion disc, 
or be ejected into tidal tails. None-the-less it was recognised that for typical properties of a merger such a
transient would yield a rapidly evolving, faint transient, significantly faster and fainter than normal core-collapse
SNe \citep[e.g.][]{li98}. Like a supernova, these events would start faint and rise to peak on timescales
of hours to days, and so would be seen as photometric bumps, interrupting the otherwise smooth
decay of the afterglow. These events have had a range of names over the years, but those which 
are commonly used today are either kilonova (referring in essence to something about 1000 times brighter than a nova), 
or macronova. 
The precise properties of these transients were uncertain, and while some
populations of faint, fast transients have been uncovered by recent synoptic sky surveys \citep[e.g.][]{kasliwal12,foley13}, none have been 
interpreted as NS-NS or NS-BH mergers (although some have been suggested to arise from NS-white dwarf mergers, through
channels similar to those creating NS-NS binaries, \citep{metzger12b,lyman14,lyman16}. 
Hence, there is little {\em observational} evidence as to the signatures that observers should search for. 
Indeed, early, deep observations of several short GRBs failed to uncover any sign of the moderate to late time ``bumps" expected
in this scenario \citep[e.g.][]{hjorth05a} and ruled out the most optimistic scenarios for these transients \citep{metzger12a}.

However, an important revelation came from detailed calculations conducted in 2013 \citep{barnes13}. It had long
been recognised that NS-NS mergers were promising sites for r-process nucleosynthesis, and that
NS-NS mergers (depending on the rates and individual r-process outputs) could be important, if not 
dominant sites for the creation of the heaviest elements \citep{rosswog99,rosswog03}. Indeed, a range
of recent studies from the abundances of radioactive elements on the sea floor \citep{wallner15} to in depth
analysis of metal poor stars in the Milky Way \citep[e.g.][]{macias16} suggest that the r-process abundances are not
in an equilibrium that would be expected if they were regularly replenished with small amounts
of additional material, as might be expected for core collapse supernovae. Instead, these analyses prefer
a scenario in which rare events, with significantly more mass per event are dominant, favouring 
a merger origin. This nucleosynthesis should naturally power a luminous transient, but because
of the dominance of r-process elements its evolution might be quite different to those 
previously assumed based on our studies of supernovae. Indeed, the opacities that were assumed in 
the earlier predictions were predominantly those of iron group elements which are dominant in normal SNe. 
The impact of including new opacities for heavy elements, in particular, lanthanides, was profound. The heavy
opacity extinguishes essentially all the optical light for external observers and means that earlier observations
were, in essence, unconstraining of the nucleosynthesis taking place. These observations instead suggested
that IR observations were the most promising route to the identification of a kilonova \citep{barnes13}. 

\begin{figure*}
  \includegraphics[width=1.10\textwidth]{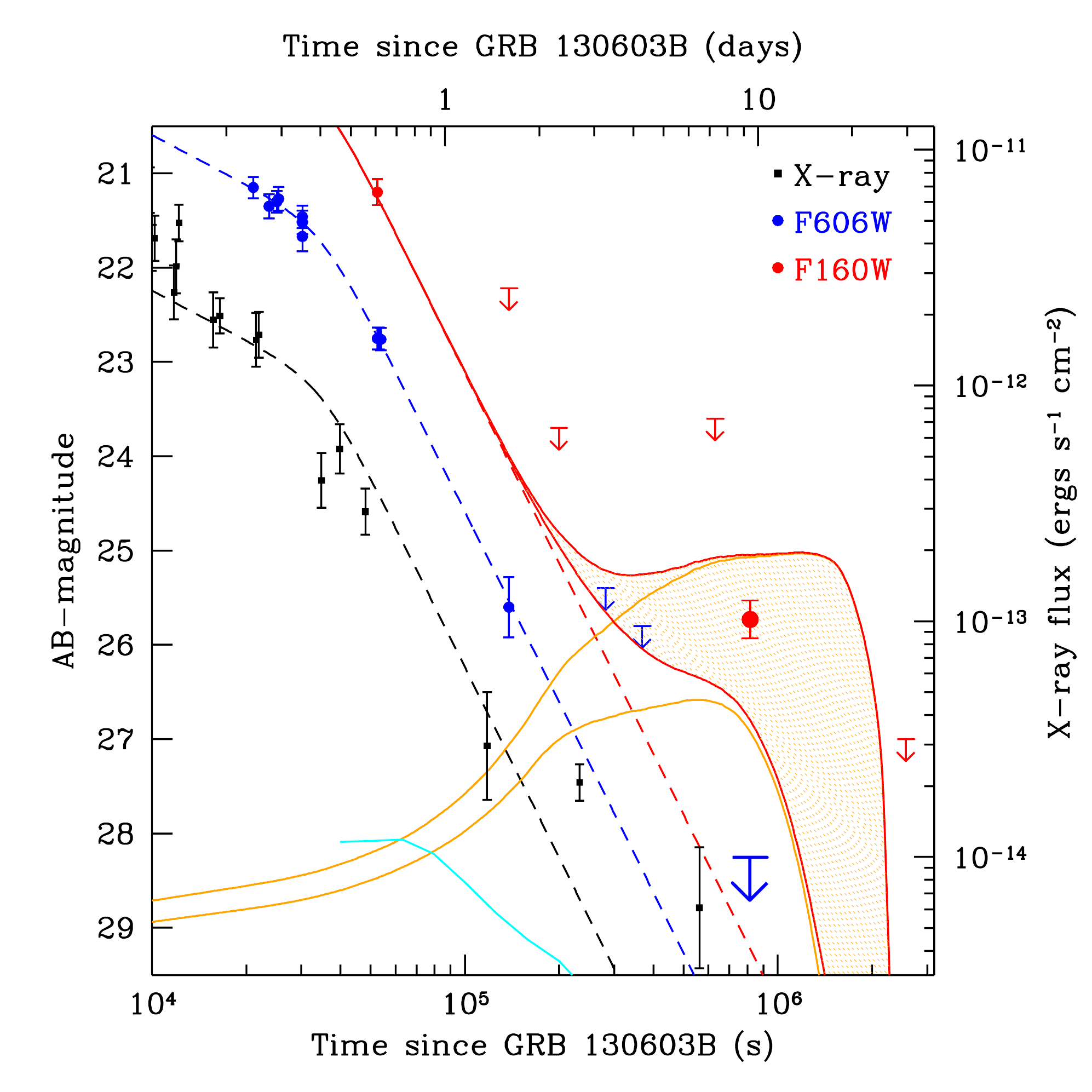}
\caption{The light curve of the kilonova seen in GRB 130603B \citep[updated from][]{tanvir13}. The points represent the X-ray
(black) optical (blue) and IR (red) photometry of the afterglow, along with their expected decay. The shaded region
is the expected kilonova range for $10^{-1}$ and $10^{-2}$ M$_{\odot}$ of material from \cite{barnes13}. While the cyan line shows
the very faint expected optical emission. The re-brightening in the IR is strongly suggestive of a kilonova, although it is also
relevant to note that late time X-ray observations by \citep{fong14} also imply that the X-ray lies above expectations. }
\label{kne}       % Give a unique label
\end{figure*}

This supposition was observationally tested with 
{\em HST} observations
of the short GRB 130603B in June 2013. These observations took place in two bands, one in the optical and
one in the IR, for a burst at $z=0.35$ \citep{deugartepostigo14}. Although only two epochs were obtained
they successfully showed a fading IR source, while nothing was seen in the optical \citep{berger13,tanvir13}. 
A comparison with the IR afterglow decay suggested a magnitude much brighter than expected, and hence
a re-brightening from an associated kilonova (Figure~\ref{kne}). 
The luminosity of the IR source was very similar to the prediction of kilonovae using the improved
lanthanide opacities, and providing the first direct evidence for the origin of short GRBs in compact
binary mergers.

Unfortunately, opportunities to further hone our understanding remain limited due to the rate of
short GRBs at sufficiently low redshift, and so to date there are limited constraints on the true properties
of this KNe. A detailed re-analysis of some archival short GRBs provides some evidence for similar
components in their light curves \citep[e.g.][]{jin15}, although other short GRBs show optical or X-ray bumps
apparently not associated with the same physical process \citep{perley09}. Indeed, recent work,
partly motivated by an apparent X-ray bump co-incident with the infrared kilonova in GRB 130603B \citep{fong14} has
focussed on possible long-lived X-ray manifestations of mergers. In particular, in the form of scattered X-ray's from the
central engine (either millisecond magnetar or black hole) \citep{kisaka15}, the absorption and re-radiation of X-rays to 
provide IR and X-ray signals \citep{kisaka16}, or direct emission of X-rays from the engine itself \citep{sun16}, perhaps creating
isotropic X-ray emission that could be of some value in searches for mergers without a GRB trigger (see below). 
However, given the paucity of observations to date we have still to distinguish between various possible suggestions for the emission processes at
play, including the role of X-ray power, the true nuclear yields and the balance between
material ejected into the accretion disc or into tidal tails around the merger.

\subsection{Gravitational Wave detection}

The mergers of compact objects were long expected to be the first observed gravitational wave signatures, and indeed NS-NS and NS-BH were 
long considered prime suspects for this. The recent discovery of a binary black hole merger with a particularly high total mass ($>60$ M$_{\odot}$) 
is therefore rather surprising at first sight \citep{abbott16a}, although it should be noted that the significantly higher masses here than normally
considered result in a far larger horizon for massive BH-BH than for NS-NS systems (the measured strain scales approximately as $M_1 M_2$), 
and so a significant astrophysical population of BH-BH binaries may make them preferentially detected \citep{abbott16c,abbott16b}. 
However, at the strain sensitivities now reached by the next generation gravitational wave detectors, and given the inferred rates
of NS-NS and NS-BH mergers from both population synthesis and observed populations (see below and \cite{abadie10}) it is
expected that mergers containing neutron stars could be found shortly. 

\begin{figure*}
\hspace{-1cm}
  \includegraphics[width=1.15\textwidth]{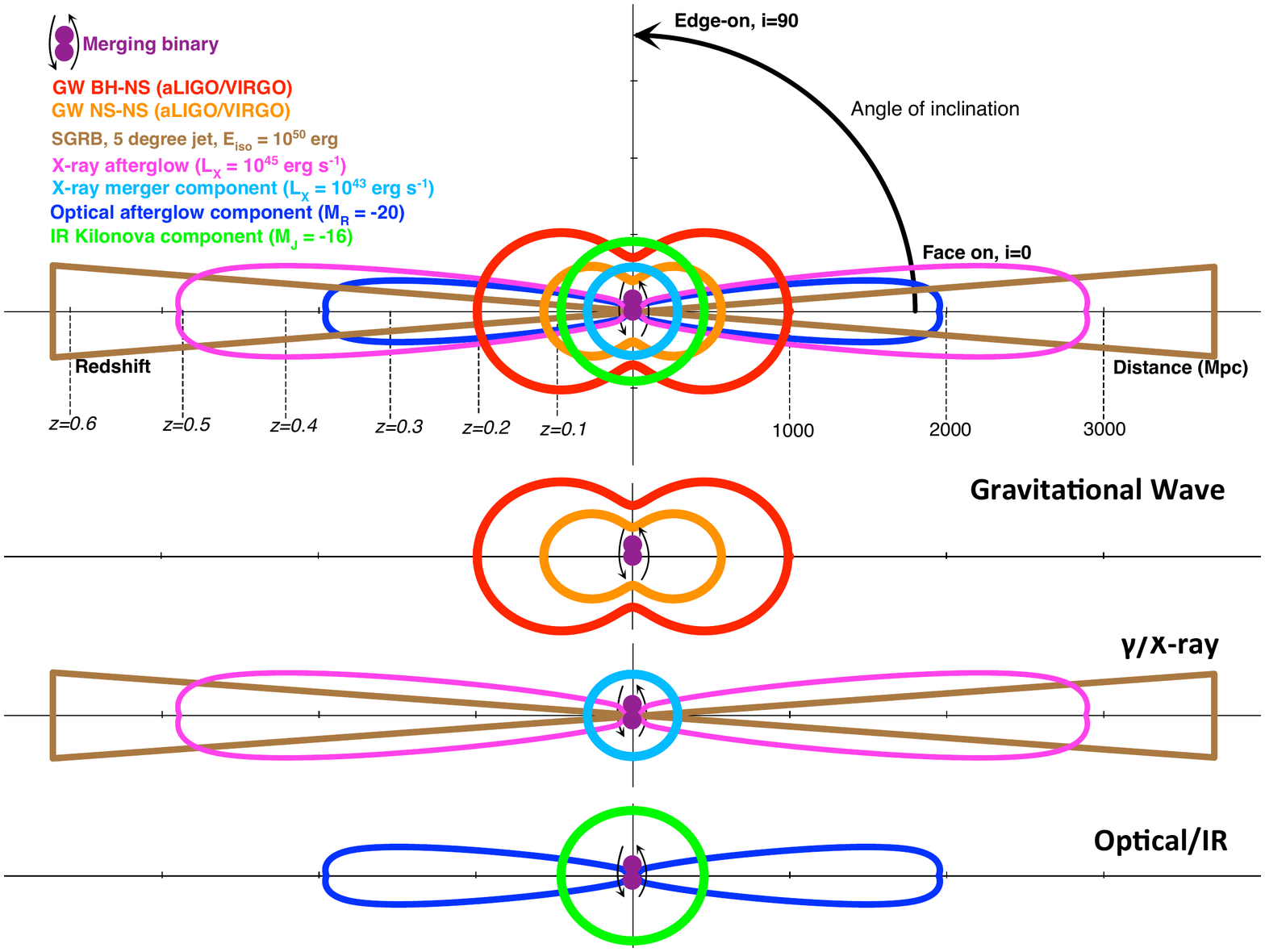}
\caption{The characteristic visibility horizons for the detection of different observational manifestations of compact object mergers. The top
panel shows all the likely signatures overplotted for comparison, while they are broken down by detection route in 
the lower panels. Mergers
can be detected by the Advanced generation of LIGO and VIRGO detectors out to hundreds of Mpc for both NS-NS and NS-BH mergers \citep{nissanke11},
although there is a strong bias towards face-on events, which can be observed to distances factors of $>2$ larger than
for edge on systems. This increases the probability of observing them in coincidence with short GRBs, which are also 
likely to be visible only to face on observers. However, such overlapping events will likely remain in a minority. 
Other wavelengths of emission are likely to require triggered observations after the detection of a gravitational wave or
short-GRB trigger. The horizons given are based on typical depths likely to be achieved fairly short exposures, several 
hours-days after the trigger with the {\em Swift} BAT ($S_{\gamma} = 1 \times 10^{-7}$ erg cm$^{-2}$, although the BAT
in practice triggers on short GRBs based on their peak flux), {\em Swift} XRT (limiting $F_X \sim 1 \times 10^{-12}$ erg s$^{-1}$ cm$^{-2}$ in $\sim 1000$s), ground
based robotic optical telescopes (R=21) and the VISTA survey telescope in the IR (J(AB) $\sim 22$). These horizons can 
obviously be scaled to different depths or time after trigger for a given assumption of temporal behaviour. In particular, observations
of short-GRBs require much smaller areal coverage and so probe much deeper and farther. For example, {\em HST} can
see kilonova signatures out to $z \sim 0.4$. Furthermore, uncertainty as to the detailed properties
of the emission means that any such horizons should be viewed as indicative rather than precise.  In particular, we properties of off-axis GRB emission which enables bursts to be
viewed at larger angles than a direct GRB remains to be directly tested, and the properties of isotropic optical/IR \citep{barnes13} or X-ray kilonova/macronova \citep{kisaka15,sun16} that
might be powered either directly by radioactivity, or by the action of a central engine such as a magnetar, remain uncertain. For SGRBs beamed signals have been
the dominant source of information to date, however, it seems likely that isotropic signatures will be more promising in the majority of gravitational wave
transients.}
\label{sgrbdist}       % Give a unique label
\end{figure*}

The simultaneous detection of GW and GRB signals offers significant advantages. The GRB signal provides a precise time and
location for a GW search. This in turn dramatically reduces the number of trials that must be run on the gravitational wave
interferometer data stream, and means that the effective sensitivity increases substantially, perhaps by a factor $\sim 2$ or more \citep[e.g.][]{dietz13,kelley13}.
However, these advantages also extend to the study of short GRB progenitors. 
For example, the chirp mass determined by 
the GW detection would immediately identify a system as either a NS-NS or NS-BH merger. Indeed, the combination of 
GRB, KNe, and GW detection provides a series of different routes to probing the compact binary population over very different distance scales (see Figure~\ref{sgrbdist}). 
In particular, SGRBs can be detected out to $z\sim 1$ or beyond \citep[e.g.][]{graham09,thoene11b}, but likely only illuminate a small fraction
of the sky due to their relativistic jets. This makes the probability of joint GRB-GW triggers relatively small, as the fraction of mergers
in a volume limited sample where the GRB jet is aligned with the observer is low.  Beaming is highly uncertain in short-GRBs, but beaming factors
$>10-100$ seem likely \citep[e.g.][]{chen13}. This is somewhat offset since the relativistic jets are visible to observers face on to the merging binary, 
a geometry which also maximises the detectability of the gravitational waves \citep[see Fig~\ref{sgrbdist} and][]{nissanke11}. 
The most promising counterparts are therefore systems that may emit isotropically. Kilonova are visible to sensitive IR searches over 
distances of several hundred Mpc, well matched to the sensitivities of current gravitational wave detectors. Similarly, isotropic X-rays with 
modest luminosity ($L_X \sim 10^{43}$ erg s$^{-1}$) should also be visible to {\em Swift} or other X-ray telescopes out to distances $>100$ Mpc. 
It is likely that both gravitational wave detections, and observations from the $\gamma-$ray to IR, and possibly beyond 
(for example magnetars should provide long-lived radio emission, \citep[e.g.][]{gompertz15,fong16}) will be necessary to fully constrain compact object
mergers, their rate, and their role in r-process production. 

\section{Central engines: linking the most luminous explosions}
 
While progress towards the nature of the progenitor stars of long GRBs has been impressive there has been increasing focus on the nature of
the compact object formed in the core of the collapsing star. These objects are the engines which drive the explosions and
create the ultra-relativistic jets which ultimately pierce the star. Traditional models extract the GRB energy through some form
of accretion onto a nascent black hole \citep{woosley93,fryer99}. Accretion rates of up to several solar masses per second
at typical efficiencies ($\sim 10\%$) can create the necessary luminosities to explain both long  and short GRBs \citep[e.g.][]{oechslin06}, provided that
a sufficiently low density of baryons exists, which seems plausible along the rotation axis of rapidly rotating stars. 
 Recently there has been a significant shift of focus to consider energy
input from a millisecond-magnetar, a newly formed highly magnetic neutron star with $B> 10^{14}$G, and spin periods of a few milliseconds \citep{metzger11}. 
Note that these, newly formed magnetars are quite different from the population of magnetars identified in the Milky Way as Soft Gamma Repeaters (SGRs) and
Anomalous X-ray Pulsars. The Galactic systems have typically lower fields, and importantly much slower rotation rates (several seconds) than
 those powering GRBs \citep[see e.g.,][]{magcat}. Indeed, even if the spin-down evolution of the Galactic magnetars is reversed, it seems unlikely that their early
 properties were consistent with those postulated as GRB central engines \citep{rea15}.
The strong magnetic fields
result in rapid extraction of energy from the dipole field, and the total energy extracted can approach $10^{52}$ erg, comparable to the 
typical isotropic energy releases of GRBs, and significantly above the beaming corrected values. Indeed an apparent upper limit
on the kinetic energy releases of GRB-supernovae of around $10^{52}$ erg provides additional support for such a model \citep{mazzali14}. 
These millisecond-magnetars would eventually spin down to 
much longer periods, at which point they would either become dormant or, depending on their mass may collapse to black holes at the point
at which they cease to be centrifugally supported.

In any case, it is clear that in the majority of GRBs this central engine inputs a large fraction of a solar rest mass on a timescale of tens to hundreds of seconds, 
perhaps extending to several thousand seconds in some cases. Indeed, evidence for prolonged energy injection existed before the launch of {\em Swift} 
 \citep{vaughan04,watson06}, but was only clearly identified following the direct detection of X-ray flares in the afterglows bursts early in the {\em Swift} era  \citep{burrows05,nousek06},
 and was surprisingly found in both long bursts and short bursts \citep{berger05}.
 While
in long bursts this late activity might naturally arise from processes involving fallback accretion, such
emission was not naturally expected in the case of short GRBs should they arise from neutron star mergers, since the
merger itself should be over very rapidly, with a relatively clean environment. Hence, if a common origin for the
long  and short GRBs was in accretion onto the nascent black hole we might expect the clean environment to
provide a significant observational distinction between the two scenarios. While short GRB afterglows are markedly
fainter, the presence of long-lived emission does not offer such easy solutions. This may be due to instabilities 
that build up in the accretion discs around both long  and short GRBs and deliver similar flares \citep{perna}, or
because some fraction of short GRBs arise from black hole-neutron star mergers, in which the neutron star
is tidal shredded over several pericenter passages \citep[e.g.][]{davies05}. 

However, the millisecond-magnetar model may be particularly appealing here. Millisecond-magnetars are likely created from the collapse
of some fraction of massive stars (see below) while they could also be formed via the merger of either
white dwarfs or neutron stars, provided the final mass is below the maximum mass of a neutron star \citep{usov92,levan06}. 
These neutron stars with extreme
magnetic fields can have rapid spin-down times, or may even be unstable and centrifugally supported, such
that as magnetic braking slows their rotation from $\sim 1$ms at birth they ultimately collapse to form
black holes. Importantly, millisecond-magnetars provide a route to providing energy input into the GRB afterglow
on timescales of minutes to hours (or potentially even longer) after the initial burst. While the models have
numerous free parameters, millisecond-magnetar models can explain many long  and short GRB light curves \citep[e.g.][]{rowlinson13}. However,
this injection of energy is not without its problems, since it should in many cases yield a strong radio afterglow. The absence of
such afterglows may be difficult to remedy with rapidly spinning magnetar models \citep[e.g.][]{fong16}. 

Interestingly, central engine models have become increasingly popular in explaining not only GRBs but also the most
luminous SNe. In this case, the energy extracted from the millisecond-magnetar re-energises the outgoing supernova shock and 
creates the additional luminosity, boosting the original luminosity by a factor of 100 or more \citep{kasen09}. The difference
between GRB magnetars and those postulated in the SLSNe lies in the duration of the energy input. Millisecond-magnetars driving SNe must
do much of their energy injection at late times when the supernova is large, otherwise, the energy may do work on the ejecta (increasing
the ejecta velocity) but not create luminosity. In contrast, in most GRBs, the spin-down times of the millisecond-magnetars are short, such that
most of their work is done more quickly. Broadly speaking GRB millisecond magnetars require high fields ($B \sim 10^{16}$G) and
spin down on timescales of hundreds of seconds, those powering SLSNe have more modest fields ($B \sim 10^{14-15}$ G) but
spin down scales of days to weeks \citep[e.g.][]{metzger15}. The ultra-long GRBs \citep{levan14} offer an interesting intermediate population 
in which millisecond-magnetars may be active for hours to days, but not longer. In this case the detection of a luminous (although not 
super luminous) supernova, SN2011kl associated with GRB 111209A \citep{greiner15} is of particular interest in providing 
evidence of a direct link between the progenitors of GRBs and those of SLSNe. 

While central engine models are now increasingly used to explain a variety of exotic explosions, it is important to note that
the engines themselves are not the same. Indeed, both black hole accretion and the magnetic extraction of 
rotational energy are considered as routes. The stars that create either black holes or rapidly spinning neutron stars may, in fact, be quite different immediately prior to the collapse, and so distinguishing between the different models is important, not
only in understanding the explosions but in characterising their progenitors. Such a task is not trivial since the central engine
itself is hidden at the core of the explosion. While the difference between asymmetric supernovae and 
ringing down magnetars is potentially distinguishable via gravitational wave observations \citep[see e.g.,][]{fryer01,davies02,rowlinson13},
such work likely beyond the capability of the current generation of detectors, that can only see such signatures
for very local (e.g. local group) supernovae. However, the rate of energy deposition and the total energy budget
are different between the two models. For example, 
in the millisecond-magnetar case rotational energy is released following $\dot{E_{rot}} = I \Omega \dot{\Omega}$, where
$E_{rot}$ is the rotational energy, $I$ the moment of inertia and $\Omega$ the spin frequency. In this case the total energy
budget is $E_{rot} = {1 \over 2} I \Omega^2$ \citep{lorimer04}. For a 1.4 M$_{\odot}$ neutron star, spinning at a period of a millisecond
(approximately the maximum spin rate for most neutron star equations of state) the total energy is of order $E_{rot} \sim 10^{52}$ erg. Alternatively, in the black hole case, the
late time accretion is likely to follow the fall-back rate of $t^{-5/3}$, and the total energy is governed simply by the total 
mass accreted $E _{acc} = \eta m_{acc} c^2$ or $\dot{E}_{acc} =   \eta \dot{m}_{acc} c^2$. For an efficiency of 10\% and
a massive star with $\sim 10$ M$_{\odot}$ of material in-falling (either directly, or via fallback) the 
total energy is a much larger $E_{acc} \sim 2 \times 10^{54}$ erg. Although the emission geometry can make
it difficult to measure the true total energy of a given explosion, it is apparent that the most energetic GRBs and 
supernovae appear to exceed the limit for neutron star energy \citep[e.g.][]{metzger15}, potentially posing 
a challenge to such models. However, there are some possibilities, such as very massive neutron stars \citep{metzger15},
that can provide a modest boost (factors of a few) to the total rotational energy, and so do enable such models
to remain plausible, even at the high energy end of the distribution.

\section{Massive star progenitors}\label{theory}
\subsection{The role of rotation} 

Observations make it clear that massive stars are now required in at least the vast majority, and probably all, long-duration GRBs. These massive stars must have somehow lost their hydrogen and helium envelope, create 
significant quantities of nickel, and have locations in galaxies consistent with the youngest and most massive stars. 
However, it is likely that not all stars with these conditions will launch a GRB since it would require very narrow
beaming angles in GRBs for the GRB rate to match the massive star rate e.g. 1 degree for progenitors with 
$M_{\mathrm{ZAMS}} > 40$ M$_{\odot}$ \citep{podsi04}. Indeed, the creation of a GRB is likely to require the
specific conditions that can give rise to a central engine to power the burst. In the
black hole engine model a centrifugally supported disc is formed
outside a newly created black hole, the GRB jet can then be launched either through electrodynamic
processes, or via neutrino -- antineutrino annihilation off the disc (note that the $T^2$ dependence 
of the weak interaction cross section means that at the extreme temperatures of these discs the 
cross section for neutrino interactions is significant). Alternatively, in millisecond-magnetar models, the magnetar must
be created, and have sufficient field and spin to energise the explosion. In both of these scenarios a crucial
factor in the creation of the GRB itself arises from {\em rotation}. 

In particular, the critical rotation is that of the core immediately prior to core collapse. In order to 
create a centrifugally supported accretion disc at the innermost stable orbit of a black hole 
a minimum specific angular momentum ($j)$ is required (i.e. $j$ is the angular momentum ($L$) per unit mass ($M$), so $j= L/M$); 
\begin{equation}
j > {\sqrt{6} G M \over c}.
\end{equation}
For typical core parameters this corresponds to $j> 10^{16}$ cm$^2$ s$^{-1}$. A newly formed millisecond-magnetar (radius $\sim 10-20$ km, and
spin period 1 ms) has a rather similar specific angular momentum, and so the rotation properties of the core prior to core collapse are likely
the same for the millisecond-magnetar or black hole scenarios. 

In principle, such angular momentum should be easy to attain. As the core grows (and increases in density) during the
main sequence its rotation should increase, potentially even to the point of breakup where the centrifugal force at the equator
is equal to the gravitational force. However, in practice, angular momentum is effectively transported {\em outwards} in stars, both
from the core to the envelope and subsequently from the envelope into the interstellar medium. The latter stage of this process is well
understood, since mass loss from the star (at its surface) carries away angular momentum, resulting in a star with lower specific
angular momentum. This process is strongly metallicity dependent, with the mass loss rate of
iron group elements scaling broadly as metallicity, $Z^{0.7-0.8}$ \citep{vink01, vink05}. Angular momentum transport within the star 
is less well understood, although it seems likely that magnetic torques would be created by differential rotation of the core relative to the outer layers,
and that these would ultimately create a rotation of the core that was tied to that of the envelope \citep{spruit02}. Indeed, while 
such a model is far from universally accepted, its use does provide a reasonable match to the spins observed in
neutron stars \citep{heger05,suijs08}. Since this mechanism couples the rotation of the core to that of the envelope, it therefore follows that the core is effectively braked by radiatively driven mass loss \citep{langer98}, and so even at modest
metallicity (e.g. somewhat less than solar) most massive stars will fall short of the critical specific angular 
momentum by an order of magnitude or more. Indeed, the apparent lack of angular momentum in most stars creates
a significant problem in understanding GRB creation. GRB-SNe are exclusively hydrogen poor events requiring no hydrogen envelope. 
However, standard pictures to create stars of this type remove the envelope via winds or binary interactions, 
through which significant angular momentum is also lost. At first sight, then the requirements of
hydrogen deficiency and rapid rotation cannot both easily be met. However, there are solutions that may provide the necessary conditions 
for GRB creation.

\subsection{Single star scenarios}
We can consider solutions that involve both single and binary stars. In this sense we take single stars to mean stars that have either lived
their entire lives (from zero age main sequence onwards) as single stars, or those in wide binaries in which there is no 
significant interaction of the two components\footnote{Indeed, it has been suggested that these wide binaries are actually the
best test-beds for single stars, since when observing a single star it is extremely difficult to tell if it has formerly been in a binary \citep{sana12}.}.
``Normal" massive single stars evolve to have a strong central concentration, in which material in the core is
used as the fuel for fusion of progressively heavier elements. On the main sequence, this core gradually builds from hydrogen to helium, and
the outer layers of the star remain (chemically) as they were at the beginning of the main sequence. 
However, in rapidly rotating stars the mixing timescale can be
shorter than the nuclear timescale, such that material synthesised in the core of the star is broadly mixed throughout the star, creating a
star which undergoes so-called {\em chemically homogeneous evolution} \citep[e.g.][]{yoon05,woosley06,yoon12}. 
This is possible because rotation creates hydrodynamic instabilities at the boundary between the convective core (where hydrogen fusion occurs) and
the radiative envelope of the star. 
Of particular importance is the Eddington-Sweet circulation \citep{eddington26,sweet50} which is driven by a thermal imbalance in rotating stars and 
may result in a short timescale of chemical mixing with a sufficiently high rotation rate. 
Because these stars evolve homogeneously 
they do not create the standard core-envelope structure, do not undergo giant branch phases, and do not experience the core-envelope breaking
that dramatically slows rotation in the cores of slowly rotating massive stars. 
However, this phase must be initiated early in the life of the star before a significant chemical gradient at the boundary between the hydrogen-burning convective core and the radiative envelope is built up, as otherwise, rotationally-induced mixing becomes too inefficient to  make the star undergo the chemically homogeneous evolution.
This process of chemically homogeneous evolution further
removes the hydrogen from the star (since it is burned) while not resulting in such significant mass loss, enabling relatively massive
cores to be built from more modest initial masses \citep{yoon05}. 

\begin{figure*}
  \includegraphics[width=1.0\textwidth]{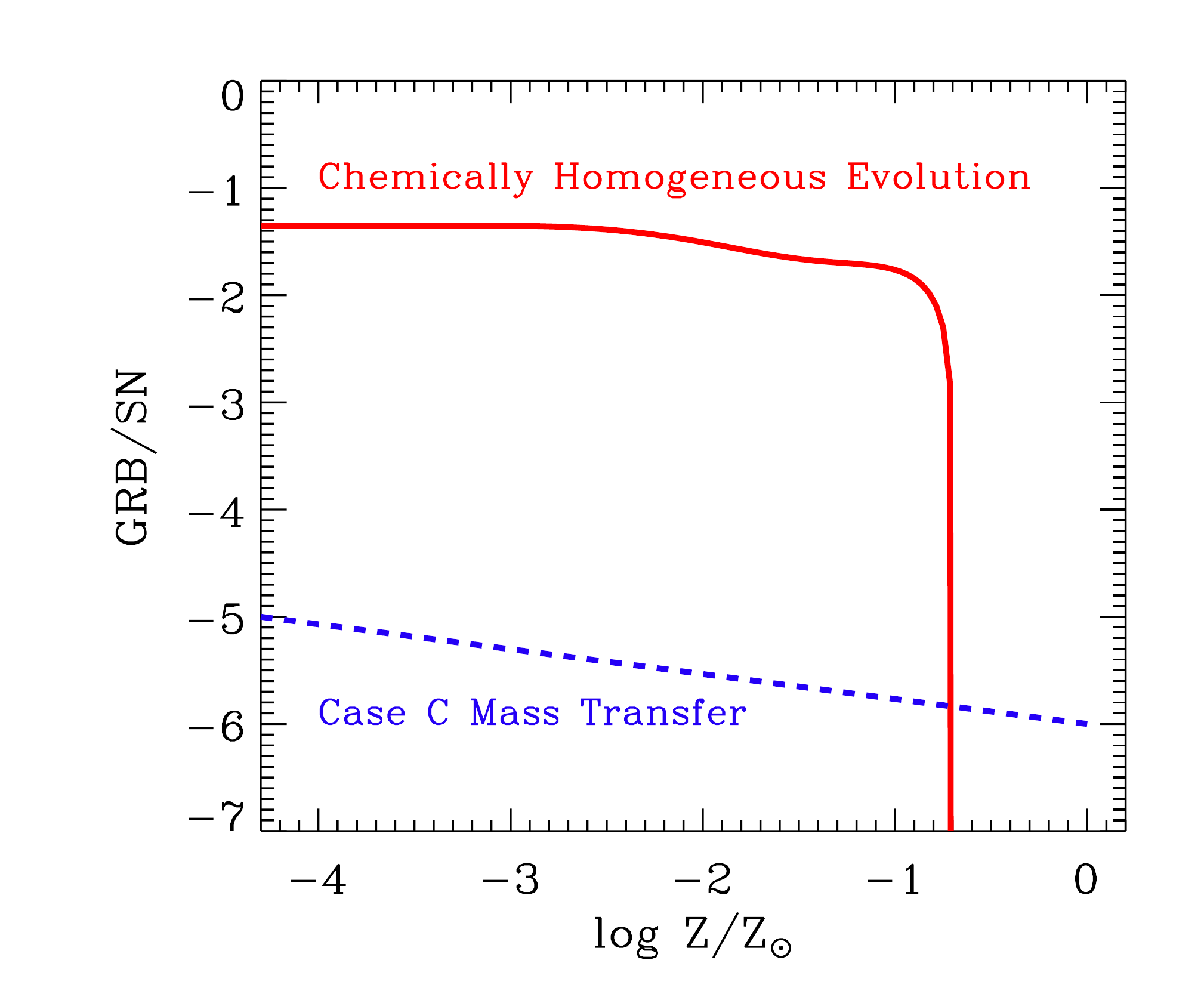}
\caption{The expected relative rates for the creation of rapidly rotating massive stellar cores through late binary mass transfer after the 
main sequence \citep[so called case C, blue line][]{podsi10}, and through chemically homogeneous evolution \citep[red line][]{yoon06}. Both are plotted as  a function of the
metallicity of the star. It is clear that at low metallicities the effectively single star rate is much larger than the Case C mass transfer
rate, and so will dominate over it as a route to GRB creation. In practice massive rapidly rotating stellar cores can be created
through routes other than case C mass transfer (mergers etc), but none-the-less it appears that a significant contribution from
homogeneously evolving stars is likely at low metallicities. Any high metallicity events would necessarily arise through binary routes. }
\label{grbrate}       % Give a unique label
\end{figure*}

However, chemically homogeneous evolution is not common. There is little evidence for it in massive stars in the Milky Way, although observations
of massive stars in the Magellanic Clouds do show some that match the expectations of chemically homogeneous evolution \citep{martins09}. This suggests that
there may well be some metallicity dependence on stars which undergo such an evolutionary pathway. Indeed, angular momentum loss
due to stellar winds is important even early in the lives of stars, so that stars have a certain threshold rotational velocity which is a function
of both their stellar mass and chemical composition. The overall impact of this is that single stars can only evolve homogeneously, and
hence create GRBs if they are of low metallicity. Indeed, the predicted rates of GRBs from chemically homogeneous stars drop rapidly at metallicity around 
0.2 Z$_{\odot}$ (see Figure~\ref{grbrate}). Interestingly, this is very similar to the critical metallicity of 0.3 Z$_{\odot}$ inferred from detailed observations of some host galaxies
by \cite{graham15b}, but somewhat lower than that inferred from larger samples (typically at higher redshift) either via spectroscopy \citep{kruhler15}
or photometric proxies \citep{perley15}. However, at these low metallicities, the contribution of these effectively single stars could be very important.

\subsection{Binary star scenarios}
Binary stars also provide a natural route of retaining or gaining angular momentum. In particular, stars may be spun up either via the accretion of
material with high specific angular momentum, via tidal locking in tight binaries, or possibly even by direct mergers. 
It is pertinent to understand binary {\it population}
properties in general, since these populations, in turn, provide direct constraints on the rate of
much rarer events, such as those giving rise to both long  and short GRBs. It is now well-established, at least
in the general field of our Milky Way galaxy and in low-density
star-forming regions, that of the most massive stars at least
60\%--80\% occur in main sequence--main sequence binary systems 
\citep[e.g][]{mason98,kouwenhoven05,raghavan10,sana10,sana11,sana12,kobulnicky14}

The highest binary fractions involving massive, and hence young, stars
are found in populous compact star clusters \citep[e.g.][and references therein]{elson98,hu10}, interestingly, the
places where the youngest most massive stars are likely to reside. This is intuitively
exemplified by the much higher fraction of low- and high-mass X-ray
binaries found in both globular clusters and actively star-forming
regions \citep[e.g.][]{coleiro13} than among the general field
star population. Indeed, interactions in the dense cores of these clusters
tend to leave the most massive stars in binaries through exchange 
interactions. These simultaneously harden tight binaries, 
at the expense of widening, and potentially unbinding the wider binaries.
This so-called hard/soft divide means that tight (hard) binaries 
get tighter, with their individual components more likely to interact and wider (soft) binaries become
progressively wider, and less likely to interact \citep[e.g.][]{heggie75}.

Given the apparent preference that GRBs have for 
low(er) metallicity environments, studies in the Magellanic Clouds may be
particularly insightful.
For example, studies of the 
population in the 15--30 Myr-old compact Large Magellanic Cloud (LMC)
star clusters NGC 1818 \citep{hu10, degrijs13,li13b},
show that for binaries with an F-star primary
and mass ratios, $q \equiv m_2/m_1 >0.4$ (where $m_1$ and
$m_2$ are the masses of the primary and secondary components,
respectively), the cluster's binary fraction is $\sim 0.35$. This
suggests a total binary fraction for F stars of 0.55 to unity,
depending on assumptions about the form of the mass-ratio distribution
at low $q$. A similarly high binary fraction 
(covering the clusters' full observable mass ranges) is obtained for the
equivalently young, compact cluster NGC 1805 \citep{li14}, as well
as for the intermediate-age ($\sim$ a few $\times 10^8$ yr-old),
massive LMC clusters NGC 1831 and NGC 1868 \citep{li14}.

However, these binary fractions represent the entire
population of binary systems. In a surprising development \cite{degrijs13}, 
noticed that the binary fraction in the
core region of NGC 1818 was significantly lower than that in the
cluster's outer regions and even in the surrounding field. It seems
likely that this is the effect of the dynamical disruption of soft binaries
on short timescales in the dense inner core of the cluster where interaction
times are low. Indeed, subsequent $N$-body simulations 
confirm this scenario \citep{geller13}.

For a range of initial conditions, from smooth virialized
density distributions to highly substructured and collapsing
configurations it is possible to explain the structures of the cluster, 
although with a slight preference for structured initial conditions. 
These models produce the observed
radial trend in binary frequency through disruption of soft binaries
(with semi-major axes, $a > 3000$ AU), on approximately a crossing
time, preferentially in the cluster core. Mass segregation
subsequently causes the remaining binaries to sink towards the core
\citep{geller15}. Thus, both a radial binary fraction
distribution that falls towards the core (as observed for NGC 1818)
and one that rises towards the core (as for dynamically older star
clusters) can arise naturally from the same evolutionary sequence
owing to binary disruption and mass segregation in rich star
clusters. Indeed, the radial distribution of the
binary fraction in  another very young LMC cluster NGC 1805,
showed an enhanced binary fraction in its core. This cluster is
dynamically much older than NGC 1818, however, so that the effects of
dynamical mass segregation had already become dominant \citep{geller15}. 
Such complex evolution within clusters means that studies of locations 
that attempt to derive the binarity of GRB progenitors (e.g. due to kicks 
to the binaries during a previous supernova explosion) are challenging since
their initial locations might be widely varied. 

Once binaries are formed, their evolution is dictated
both by the evolution of the individual components, but crucially, also
by the interaction between the two stars, which can have a marked
impact on their evolutionary pathways. The introduction of 
binary channels greatly increases the diversity of stellar evolution pathways and
their associated remnants. Indeed, it is visible on galaxy-wide scales where,
for example, the rejuvenation of older stars via binary interactions can 
make them look much younger than they (or the bulk of the stellar population)
really are \citep{eldridge09,stanway16}. 

Because of the increased diversity of stellar products that can be created through binary evolution 
(e.g. unusual composition, rotation, mass etc), binary channels are popular routes
to create exotic (and rare) stellar transients. Through binary evolution it is straightforward to 
remove the stellar envelope (and provide a stripped envelope core collapse supernova), while tidal
locking in tight binaries can in principle provide the necessary angular momentum in the
core to enable disc or millisecond-magnetar formation. However, the impacts of binary evolution go beyond
the straightforward changes on the outer layers of the star due to the impact of centrifugal force and rotationally-induced chemical mixing. Hence, detailed studies
require the construction of complex models that can simultaneously model angular momentum
transport within individual stars and the binary components, as well as tracking mass loss and nuclear
burning. Such calculations have been attempted by several teams \citep[e.g.][]{yoon10}. 

As with single stars, the crucial parameter to track through binary routes is the specific angular momentum of the
stellar core immediately prior to core collapse. Stars can be spun up via mass transfer from a companion star, but 
are spun down if they themselves lose mass. This creates similar problems to those seen in massive
single stars, in which it is very difficult to both lose the hydrogen envelope and retain both sufficient mass and rotation
to obtain the necessary conditions for GRB creation. Several authors have considered late time 
mass transfer during a late stage of core helium burning or after core helium exhaustion (so-called Case C mass transfer) as 
a possible route to obtaining GRB-like conditions. This late time mass transfer can enable the formation of a massive
stellar core, even at high metallicity \citep{brown2004}. Alternatively, depending on the properties of the binary this mass
transfer may result in the formation of a common envelope, which will unbind the envelope of the star at the cost of
the orbital energy (and hence separation). The resulting binary (after removal of the envelope) is a tight binary containing
a carbon-oxygen core. Tidal locking within this binary may match the rotation period of the core to the period of the binary
(typically a few hours) in which case the requirements on specific angular momentum can be met \citep[e.g.][]{lee02}. 
Indeed, tidal locking through a variety of routes including neutron stars, post main sequence stars and low mass main sequence 
stars has been considered as a route of creating GRBs \citep{izzard,vdh07}, although detailed calculations 
disfavour this scenario \citep{detmers08}. 
Alternatively,
evolution in binaries with relatively extreme mass ratios may also create the necessary conditions. Here
mass transfer occurs from a massive star evolving away from the main
sequence, and onto a low mass star on the main sequence. In this scenario, a further mass transfer from the low mass
companion can occur on the CO core of the primary star. This, in turn, leads to both the spin up of the core, and the
explosive ejection of the common envelope \citep{podsi10}. 

Finally, a route to obtaining the necessary angular momentum may be the direct merger of two stars. Such mergers do happen, for example, common envelope mergers 
have been used to explain unusual stellar outbursts of lower mass stars, such as V838 Mon, V1309 Sco
or R136a1\citep[e.g.][]{tylenda06,tylenda11,banerjee12}. More massive mergers may create the elusive Thorne--Zytkov objects.
One of these has recently been claimed in the SMC \citep{levesque14}, although this is controversial, in particular, it seems 
unclear if it actually resides in the SMC \citep{worley16}, or is a foreground source \citep{maccarone16}.
 Indeed, at higher masses the merger of either two Helium cores, or perhaps a black hole (or
neutron star) with a Helium core has been proposed as a GRB mechanism, with the latter suggested as a possible
origin for the Christmas-day burst, GRB 101225A \citep[e.g.][]{thoene11}. In these mergers the orbital angular momentum
eventually is combined within the single merged object. In the case of He-star -- He-star mergers this dramatically
increases the total mass, and there is little time between the merger and the supernova. For black hole--He star mergers, the
black hole accretion can essentially create the GRB immediately. 

It is clear that there are a wide variety of binary channels that can give rise to the necessary conditions for GRB
production. All of these scenarios are likely to occur in nature. However, it is less clear if the most exotic 
events can create GRBs at the necessary rate to explain the observed populations of GRBs. 
In each of these scenarios, there is a limited parameter space over
which the channel will work (in terms of mass, mass ratio, and initial separation), and so a given route is likely to 
provide only a modest rate of GRB-like events (see Figure~\ref{grbrate}). Hence, while many stars are in binaries,  the
rates of creation of rapidly rotating massive cores through any one of these routes is significantly lower than the rate
obtained through chemically homogeneous evolution for any star formation occurring at substantially sub-solar metallicity. 

\subsection{Predicted long GRB rates}
Determining the rates of GRBs, and of their various progenitor channels is fraught with difficulty. Observationally, GRB
detection is a sensitive function of both the properties of the detector (area, energy range etc) and of the
burst (hardness, peak flux, total fluence etc). Furthermore, GRBs exhibit a broad range of luminosities, and so an 
extrapolation to the total rate of GRB-like events requires a significant correction factor. Indeed, it is apparent that there
is a population of local, low-luminosity GRBs whose volumetric rate exceeds those of the more distant bursts by a factor
of several hundred. If these are a separate population, or the faint end of a luminosity function remains uncertain \citep[e.g.][]{chapman07,liang07}, although
the similarities in the supernovae in both the lowest and highest luminosity examples \citep[e.g.][]{galama98,pian06,xu13,levan14b} 
suggests that similar physical mechanisms are at play. Finally, GRBs are relativistically beamed, and illuminate only a small fraction 
of the sky \citep{frail01,bloom03}. The beaming fraction for a given burst can be obtained from the so-called jet break when 
the relativistic jet expands laterally \citep{frail01}, but is a challenging and still highly uncertain parameter that significantly impacts
the observed rates. 

Similarly, when determining the likely rates of GRB production via the various routes considered above there are major uncertainties that
enter. What is the relevant metallicity distribution for stars throughout the Universe, and how does this impact those that will undergo 
chemically homogeneous evolution \citep{langer06}? Are there any special environments in which the apparently universal top end of the initial mass function
can become more top heavy \citep{bastian10}? 
What is the binary fraction, and the range of initial separations? How is the mass ratio of binaries distributed, are these universal, and
might this have any impact on the final products of stellar evolution \citep{li13a}? What is the efficiency
of common envelope evolution \citep[e.g.][]{nelemans05}? While these parameters can be studied from detailed simulations (or observations) \cite[e.g.][]{yoon06}, or explored via
rapid population synthesis \citep[e.g.][]{izzard} there are still order of magnitude uncertainties in the true rate. \cite{podsi04} attempted to compare the
rates of GRBs and massive stars by directly comparing the rates of GRBs (for some assumptions about beaming angles) with the rates of stars of
a given mass within a typical galaxy. In table~1 we provide an updated version of this table, including estimates of the low- and high- luminosity
GRBs separately, as well as different types of SNe and massive stars. We also provide the rates of massive stars for a typical (i.e. Milky Way-like) galaxy as
well as the rates of massive stars below some threshold metallicity in the local Universe \citep[extrapolated from][]{graham15c}. We note that these scaled values
assume a constant factor of $10^7$ between the volume averaged rate and the galactic rate, and so does not take into account chemical differences 
between galaxies of different masses.
 
The rate of classical, high luminosity long GRBs, for a beaming angle of a few degrees is remarkably similar to the rate of low metallicity $>40$ M$_{\odot}$ stars
observed in the local Universe. Since both low metallicity \citep{graham15b,perley15} and high mass \citep{larsson07,raskin08} have been observationally linked
to GRBs this may suggest that a reasonable fraction of initially very massive stars create long GRBs. Although such interpretation is challenging because
many high-luminosity long GRBs are at much higher redshift, where star formation rate densities are higher. None-the-less, this suggests that 
the rate of production of long GRBs is not much lower than the rate of production of massive ($>40$ M$_{\odot}$), low metallicity ($<0.25$ Z$_{\odot}$) stars. In turn, this implies that they are probably not created only from stars which undergo very rare and unusual interactions. For example, for a typical 5-degree beaming
angle, the LGRB rate is only a factor of 3 lower than the low metallicity massive star rate. While alternative (wider) beaming angles could lower the rate of
GRBs by an order of magnitude or more, it still seems unlikely that the rarest channels, involving $<1\%$ of stars are likely to be creating long GRBs. 
Indeed, if the low-luminosity GRBs are arising from a similar population, with similar progenitor masses and low metallicities then they would need to be born
from a significant fraction of the very massive stars, even in the case of no beaming. Even with no beaming correction, the rate per galaxy of low-luminosity 
GRBs is $\sim 2 \times 10^{-5}$ galaxy$^{-1}$ yr$^{-1}$, again only a factor of a few lower than the low metallicity 40 $M_{\odot}$ stars. Indeed, since these low-luminosity GRBs are at low redshift there is less concern about
evolution over cosmic history, and it seems likely that the low-luminosity GRBs are either lower-mass stars or nearly isotropic in emission in order to avoid low luminosity
GRB rates that approach, or even exceed the massive star formation rate. 
It is also relevant to compare the rates to those of other transients, for example, the newly uncovered population of ULGRBs \citep{levan14} that may be
linked to the SLSNe \citep{greiner15}. The beaming corrections in this case are particularly uncertain, but for plausible beaming rates, comparable to
those of long GRBs the inferred rate of ULGRBs is lower, of the same order of magnitude as the SLSNe rate. However, given the substantial uncertainties
in both rate calculations, it is possible that this similarity is purely coincidental.

\begin{table}[htdp]
\begin{center}
\begin{tabular}{llll}
\hline
Object & Rate & Rate & References \\
& (galaxy$^{-1}$ yr$^{-1}$)  & (Gpc$^{-3}$ yr$^{-1}$ )  \\
\hline
{\bf Transients} \\
\hline
LGRB  & $8 \times 10^{-8}$ & 0.8 & \cite{sun15}\\ 
 -- 1$^{\circ}$ & $6.6 \times 10^{-4}$ & 6600  \\ 
  -- 5$^{\circ}$ &$2.6 \times 10^{-5}$ & 260 \\ 
   -- 20$^{\circ}$ & $1.6 \times 10^{-6}$ & 16  \\ 
LLGRB & $1.6 \times 10^{-5}$& 160 &  \cite{sun15} \\
 -- 1$^{\circ}$ & 0.11 & $1.1 \times 10^{6}$ \\ 
  -- 5$^{\circ}$ & $4.2 \times 10^{-3}$ & 42000  \\ 
   -- 20$^{\circ}$ & $2.6 \times 10^{-4}$ & 2600 \\ 
SGRB & $2.0 \times 10^{-7}$ & 2.0 &  \cite{sun15} \\
 -- 1$^{\circ}$ & $1.3 \times 10^{-3}$ & 13000 \\ 
  -- 5$^{\circ}$ & $5.3 \times 10^{-5}$ & 530  \\ 
   -- 20$^{\circ}$ & $3.0 \times 10^{-6}$ & 30 \\ 
ULGRB & $1 \times 10^{-8}$ & 0.1 & \cite{gendre13,prajs16} \\
 -- 1$^{\circ}$ & $6.6 \times 10^{-5}$ & 660  \\ 
  -- 5$^{\circ}$ & $2.5 \times 10^{-6}$ & 25 \\ 
   -- 20$^{\circ}$ & $2.0 \times 10^{-7}$ & 2.0 \\ 
rTDE  & $3.0 \times 10^{-9}$ & 0.03 & \cite{sun15} \\
SLSNe* & $3.0 \times 10^{-6}$ & 30  & \cite{quimby13,prajs16} \\
\hline
{\bf Massive stars}  \\
\hline
20 M$_{\odot}$ all $Z$  & 2 $\times 10^{-3}$ & 20000 & \cite{podsi04} \\
20 M$_{\odot}$ $Z < 1/4 Z_{\odot}$  &  2 $\times 10^{-4}$ & 2000 & \cite{graham15c} \\
40 M$_{\odot}$ all $Z$ & 6 $\times 10^{-4}$ & 6000  \\
40 M$_{\odot}$ $Z < 1/4 Z_{\odot}$ &  6 $\times 10^{-5}$ & 600  \\
80 M$_{\odot}$ all $Z$ & 2 $\times 10^{-4}$  & 2000 \\
80 M$_{\odot}$ $Z < 1/4 Z_{\odot}$ & 2 $\times 10^{-5}$  & 200 \\
\hline
{\bf Compact binaries}  \\
\hline
NS-NS & $6 \times 10^{-5}$ & 600 & \cite{abadie10}   \\
NS-BH & $2 \times 10^{-7}$  & 2 \\
BH-BH & $5 \times 10^{-7}$   & 5 & \cite{abbott16c} \\ 
\hline
\end{tabular}
\caption{Approximate rates of various engine driven transients, and of massive stars at all metallicities, and below a set metallicity. Various attempts have been made to determine these
rates, and the numbers given are rounded and approximate, rather than representing the full range of possibilities (which in some cases are quite large). A typical galaxy is assumed to have a B-band luminosity of $10^{10}$ L$_{\odot}$, and we assume a local B-band luminosity density of $10^{8}$ Mpc$^{-3}$ \citep[e.g.][]{calura03}, although clearly, this
value evolves significantly over cosmic history. We note that the numbers per galaxy have been directly scaled from the volume averaged values with a fixed scaling
of $10^7$. Since stars of a given chemical makeup are not the same ``per galaxy" due to the mass-metallicity relation \citep{tremonti04}, the relative numbers of 
low metallicity stars should be viewed with caution since a given galaxy might contain stars entirely of low metallicity (if it was of low mass), or very few stars of
low metallicity (if it was of high mass). i.e. If low metallicity is a requirement for GRB production then the rate per galaxy at low metallicity is boosted in low mass/metallicity galaxies, and depressed in high mass/metallicity galaxies.
The massive star corrections to low metallicity (formally $12 + \log(\mathrm{O/H}) < 8.4$ following \cite{graham15c}) are at $z \sim 0$ where
10\% of the star formation lies below this metallicity. This increases rapidly, and the correction likely becomes less than a factor 2 by $z\sim 2$. . }
\end{center}
\label{grb_rates}
\end{table}%

\section{Constraints from local stellar populations}

\subsection{O stars}

It is apparent that long GRBs arise primarily from a subset of massive ($M_{\mathrm{ZAMS}} \geq 40 M_{\odot}$), moderately metal-poor stars at cosmological redshifts. Although the nearest long GRBs lie at distances of tens to hundreds of megaparsec, we can study individual massive stars within appropriate environments in the Local Group. 

The overwhelming majority of stars are believed to have their origins in dense star clusters, intermediate density OB associations or low-density star forming regions. If stars are randomly drawn from a \cite{kroupa01} Initial Mass Function (IMF), cluster masses over $\sim 100 M_{\odot}$ are required to produce one star capable of ending its life as a core-collapse supernova. However, the most massive stars are preferentially found in the more massive clusters. In particular, stochastic sampling of the initial mass function 
means that the probability of a low mass cluster forming a very massive star is low (i.e. the massive star content of say 10, 100 M$_{\odot}$ associations is less than that of a single 
1000 M$_{\odot}$ cluster). Indeed, it has been suggested that the maximum mass of a star in a cluster ($M_{*,max}$) is related to the cluster total mass $(M_C)$, such that $M_{*,max} \approx 0.39 M_C^{2/3}$ \citep{weidner10}. In this case,
for a star forming region to include at least one star with an initial mass of at least $\sim 40 M_{\odot}$, a total mass 10$^{3} M_{\odot}$ is required, with 10 such stars hosted by 10$^{4} M_{\odot}$ clusters. 

The upper limit to star cluster mass is a sensitive function of how vigorously stars are being formed, so one would not expect long GRBs to occur in galaxies with low specific star formation rates. Therefore if progenitors of long GRBs arise from very massive stars, they will exclusively occur in galaxies with high specific star formation rates. Indeed, \citet{kelly12} conclude that low $z$ broad-lined (BL) SNe Ic arise exclusively from hosts with the highest star-forming intensities, with \citet{kelly08} having earlier established that long GRBs and low $z$  SN Ic-BL originate from similar environments.

Since long GRBs strongly favour host galaxies with metallicities below 1/2 solar, the focus of our attention in the Local Group will be primarily the Large and Small Magellanic Clouds, with 1/2 and 1/5 solar metallicity, respectively. The most massive young star-forming region in the SMC, N66/NGC 346, hosts several tens of O stars, whereas many hundreds of O stars are known in the Tarantula Nebula region of the LMC. 

The VLT FLAMES Tarantula Survey \citep[VFTS][]{evans11} has enabled properties of 800 OB stars in this region to be determined, revealing that 50\% of O stars are affected by binary evolution \citep{sana13}, with relatively low rotation rates, aside from a few very fast rotators presumably arising from binary interactions \citep{ramirez-agudelo15}. Several star clusters are located within the Tarantula Nebula, most noticeably R136, the youngest (1.5 Myr) very high-mass cluster within the Local Group. \citet{crowther16} exploit HST/STIS spectroscopy to reveal that over three dozen stars more massive than 40 $M_{\odot}$ are located within the central parsec. 

Nearly 30 very massive stars (VMS, $\geq 100 M_{\odot}$) are located in the Tarantula Nebula, the majority within the R136 region, but others up to 100 parsecs away \citep{crowther16}. Some of the more remote VMS are plausible runaways from R136, but others likely were born in significantly low-density regions. In contrast, no VMS are located in SMC's N66 star-forming region, although it hosts a triple high mass system HD 5980.

In summary, large numbers of candidate long GRB progenitors exist in the LMC/SMC as far as their initial masses are concerned, yet there is no evidence that these are born with sufficiently high rotational rates for chemically homogeneous evolution. Rapid rotation ($v \sin i >$ 500 km\,s$^{-1}$)  is observed for small numbers of O stars in the Tarantula Nebula, presumably spun-up via binary evolution \citep{demink13}. Similarly, no examples of rapid rotators are observed in N66 \citep{mokiem06}. \citet{lamb16} identify a large population of SMC field Oe stars, presumably arising from rapid rotation, although these too may result from close binary evolution. This in itself raises interesting questions, since if a significant population of sufficiently massive and low metallicity stars
can be identified in the local Universe, but essentially none of these appear as viable GRB progenitors then the fraction of massive low metallicity stars creating GRBs must be small. However, 
this begins to create tension with the necessary observed rates, especially when lower luminosity GRBs are considered (see section 5.4).

\subsection{Wolf-Rayet stars}

The immediate progenitors of long GRBs are massive stars whose hydrogen-rich envelopes have been stripped away to reveal compact cores, i.e. classical Wolf-Rayet (WR) stars \citep{crowther07}. These stars possess dense, fast  outflows with atmospheric compositions characteristic of hydrogen-burning (WN-type) or helium-burning (WC-type). In common with their O type progenitors, the strength of WR winds scales with metallicity \citep{vink05}, so metal-poor WR stars possess, lower density, slower outflows than their Milky Way counterparts \citep{hainich15}.

The observed association between long GRBs and SN Ic-BL suggests the presence of very little helium in the progenitor star, favouring WC stars. The Magellanic Clouds host approximately 150 WR stars, of which only 15\% belong to the carbon sequence or rare oxygen sequence (WO).  The binary frequency of Magellanic Cloud WR stars is approximately 40\%, similar to that observed in the Milky Way, with a lower binary fraction amongst WC and WO stars \citep{bartzakos01}. 

Analysis of single WC and WO stars in the LMC \citep{crowther02, tramper15} reveals He core masses of 10--20 $M_{\odot}$. The sole SMC carbon/oxygen sequence WR star is in a close binary system, although \citet{shenar16} argue that the binary channel does not dominate the formation of WR stars at this metallicity.

The dense outflows from WR stars prevent direct rotational velocity measurements, but searches for non-spherical geometries  can be done via linear spectropolarimetry. None of the four LMC WC stars observed by \citet{vink07} exhibited the characteristic line depolarization with respect to the continuum, ruling out rapid rotation in these cases. \citet{martins09} have claimed that at least one of the hydrogen-rich WN stars in the SMC is the product of chemically homogeneous evolution. However, this single star might represent the product of spin-up arising from close binary evolution since such high initial rotation rates are not currently established amongst massive O stars in the SMC.

Finally, studies have been undertaken investigating the locations of WR stars relative to their host light, in principle directly comparable to the locations of transient phenomena
\citep{leloudas10,bibby12}. These broadly show the locations of stripped envelope SNe match those of the WR population \citep{bibby12}, and that WC stars are
more concentrated on their host light than WN system \citep{leloudas10}. 

In summary, there are no robust long GRB progenitors amongst the Magellanic Cloud WR population at present, although at least one SMC WR star has been claimed to be the result of chemically homogeneous evolution, and their weak winds make measurements of their rotational rates via  spectropolarimetry extremely challenging.

\section{Summary and open questions}

We are now approaching 20 years since the discovery of the first GRB afterglows, and to date over 1000 have been discovered. Intensive 
observations of the bursts themselves, their afterglows, associated supernovae and host galaxies have provided firm links to their progenitors
in several cases, and in particular the link between long GRBs and type Ic supernovae appears secure. In turn, the link to young massive
stars, coupled with their extreme luminosity makes GRBs a powerful probe of the distant Universe from the discovery of some of the
most distant galaxies, to detailed work studying the buildup of stellar mass and metals across cosmic history. However, while this
utility is clear, there remain many questions in addressing the nature of the progenitors themselves, and this, in turn, has a systematic impact
on the cosmological utility of GRBs (e.g. in determining the correct factor to convert a GRB rate to a star formation rate at a given redshift/metallicity). 
In this review, we have highlighted much of the progress that has been made through a combination of intensive observation, large-scale numerical 
calculation and direct observation of local analogue populations. However, there are many questions that remain open, and will be the focus
of research in the coming years, in particular;

\begin{itemize}

\item
What fraction of massive O-stars are really required to create GRBs? What does this mean about routes to their progenitors? 

\item
What is the true metallicity dependence for GRB production? 

\item
How important is binary evolution in the creation of GRBs? 

\item
Are any of the locally observed massive, rapidly rotating stars actually likely to create a GRB? Have any in the past that we can identify? 

\item
What is the diversity of supernovae seen in long GRBs? Can they be standard candles? Is the presence of a luminous SN in one case a common or 
extremely rare occurrence? 

\item
What is the range of kilonova/macronova properties seen in short GRBs? Can these create an important contribution to r-process enrichment, and
explain the levels seen in the sea-floor on Earth? 

\item
What is the role of NS-NS and NS-BH mergers in short GRB production? Can any BH-BH mergers make GRBs? What will simultaneous gravitational 
wave signatures tell us? 
\end{itemize}

The answers to these questions are likely to come both from the application of the traditional techniques described above, with improved 
telescope aperture and response (e.g. JWST, E-ELT, GMRT etc), or the every increasing computational power, and through new routes,
such as direct multi-messenger observations (gravitational waves, neutrinos) that are now producing astrophysical detections. The answers to these
questions are challenging, but it is likely that many will be uncovered with a degree of certainty in the coming decade.

\section*{Acknowledgements}
We thank the organisers and ISSI Beijing for a productive and well organised meeting in April 2015, and the referee for a very constructive and valuable report that
has improved this manuscript. 
AJL acknowledges support from the Science and Technology Facilities Council under programme ST/L000733/1 and
from the Philip Leverhulme Trust via a Leverhulme Prize. He is grateful for productive conversations with Nial Tanvir.
RdG acknowledges financial support from the National Natural Science Foundation of China (grants 11373010, 11633005 and U1631102). He also acknowledges helpful
insights into the evolution of binary systems provided by the
participants of the 2014 conference on `Binary systems, their
evolution and environments' (particularly Philipp Podsiadlowski, Chris
Belczy\'nski and Dany Vanbeveren). SCY was supported by the Basic Science Research (2013R1A1A2061842) program through the National Research Foundation of Korea (NRF).

%\begin{figure}
%  \includegraphics{example.eps}
%\end{figure}

%\begin{table}
%\begin{tabular}{lll}
%\hline\noalign{\smallskip}
%\end{tabular}
%\end{table}

%\begin{acknowledgements}
%\end{acknowledgements}
\bibliographystyle{aps-nameyear}      % American Physical Society (APS) style, author-year citations
%\bibliography{example}                % name your BibTeX data base

           % name your BibTeX data base
%\nocite{*}

%\begin{thebibliography}{}
%\bibitem[\protect\citeauthoryear{Aamport}{1986}]{RefJ}
%L.A. Aamport, \mbox{G-Animal's} Journal \textbf{41} (7), 73 (1986).
%This is a full ARTICLE entry

%\bibitem[\protect\citeauthoryear{Knuth}{1981}]{book-full}
%D.E. Knuth, \textit{Seminumerical algorithms}, 2nd edn.
%The Art of Computer Programming,
%vol. 2 (Addison-Wesley, Reading, 1981).
%This is a full BOOK entry

% Format for proceedings
%\bibitem[\protect\citeauthoryear{Oz and Yannakakis}{1983}]{RefB}
%W.V. Oz, M. Yannakakis (eds.),
%in \textit{All ACM Conferences} (Academic Press, Boston, 1983).
%This is a full PROCEEDINGS entry
% Other formats available: INPROCEEDINGS, PHDTHESIS, TECHREPORT, 
% UNPUBLISHED, MISC, MASTERSTHESIS, MANUAL, INCOLLECTION, BOOKLET
% etc
%\end{thebibliography}

\end{document}